\documentclass[a4paper,fleqn,usenatbib]{mnras}

\usepackage{savesym} 
\usepackage{amsmath}
\savesymbol{iint}
\savesymbol{iiint}
\usepackage{txfonts}
\restoresymbol{TXF}{iint}
\restoresymbol{TXF}{iiint}

\usepackage[T1]{fontenc}
\usepackage{ae,aecompl}

\usepackage{graphicx}	
\usepackage{amsmath}	
\usepackage{amssymb}	
\usepackage{aas_macros}

\newcommand{\parcsec}{\ensuremath{\,.\!\!\arcsec}}
\newcommand{\kms}{km\,s$^{-1}$}

\title[SN~2015bh]{Dead or Alive? Long-term evolution of SN~2015bh (SNhunt275)}

\author[Elias-Rosa et al.]{N. Elias-Rosa$^{1}$\thanks{E-mail: nancy.elias@oapd.inaf.it}, 
A. Pastorello$^{1}$, S. Benetti$^{1}$, E. Cappellaro$^{1}$, S. Taubenberger$^{2,3}$, 
\newauthor
G. Terreran$^{1,4}$, M. Fraser$^{5}$, P. J. Brown$^{6}$, L. Tartaglia$^{1}$, A. Morales-Garoffolo$^{7}$, 
\newauthor
J. Harmanen$^{8}$, N.~D. Richardson$^{9}$, \'E. Artigau$^{10}$, L. Tomasella$^{1}$, R. Margutti$^{11,12}$,  
\newauthor
S. J. Smartt$^{4}$, M. Dennefeld$^{13}$, M. Turatto$^{1}$, G. C. Anupama$^{14}$, R. Arbour$^{15}$, 
\newauthor
M. Berton$^{16}$, K.~S. Bjorkman$^{9}$, T. Boles$^{17}$, F. Briganti$^{18,19}$, R. Chornock$^{20}$, 
\newauthor
F. Ciabattari$^{21}$, G. Cortini$^{22}$, A. Dimai$^{23}$, C.~J. Gerhartz$^{9}$, K. Itagaki$^{24}$, R. Kotak$^{4}$, 
\newauthor
R. Mancini$^{18,19}$, F. Martinelli$^{18,19}$, D. Milisavljevic$^{25}$, K. Misra$^{26}$, P. Ochner$^{1}$, 
\newauthor
D. Patnaude$^{25}$, J. Polshaw$^{4}$, D. K. Sahu$^{14}$,  S. Zaggia$^{1}$
\newauthor
{\small \it Affiliations are listed at the end of the paper}
}

\date{Accepted XXX. Received YYY; in original form ZZZ}

\pubyear{2016}

\begin{document}
\label{firstpage}
\pagerange{\pageref{firstpage}--\pageref{lastpage}}
\maketitle

\newpage
\newpage
\begin{abstract}
Supernova (SN)~2015bh (or SNhunt275) was discovered in NGC~2770 on 2015 February with an absolute magnitude of M$_r \sim -$13.4 mag, and was initially classified as a SN impostor. Here we present the photometric and spectroscopic evolution of SN~2015bh from discovery to late phases ($\sim$ 1 yr after). In addition, we inspect archival images of the host galaxy up to $\sim$ 21 yr before discovery, finding a burst $\sim$ 1 yr before discovery, and further signatures of stellar instability until late 2014. Later on, the luminosity of the transient slowly increases, and a broad light curve peak is reached after about three months. We propose that the transient discovered in early 2015 could be a core-collapse SN explosion.  The pre-SN luminosity variability history, the long-lasting rise and faintness first light curve peak suggests that the progenitor was a very massive, unstable and blue star, which exploded as a faint SN because of severe fallback of material. Later on, the object experiences a sudden brightening of 3 mag, which results from the interaction of the SN ejecta with circumstellar material formed through repeated past mass-loss events. Spectroscopic signatures of interaction are however visible at all epochs. A similar chain of events was previously proposed for the similar interacting SN~2009ip. 

\end{abstract}




\section{Introduction}

Massive stars are known to lose mass via steady state winds or through dramatic eruptions in which they increase significantly their brightness, becoming intermediate-luminosity optical transients. In some cases, these non-terminal outbursts compete in luminosity with real supernovae (SNe), and may also mimic their observables. For this reason, they are commonly known as ``SN impostors'' (e.g. \citealt{vandyk00}). As a consequence, these luminous eruptions of massive stars may be misclassified as genuine SNe. This is what frequently happens with giant eruptions of massive stars such as luminous blue variable (LBV) stars, whose spectra are characterised by incipient narrow (full-width-at-half-maximum - FWHM - lower than about 1000 \kms) hydrogen lines in emission, resembling those observed in Type IIn SNe. In Type IIn SNe, the narrow features are usually interpreted as signatures of interaction between the SN ejecta and the circumstellar medium (CSM) embedding the SN. In general, the discrimination between SN impostors and Type IIn SNe is often controversial (see e.g. SN~2011ht-like objects, \citealt{roming12} and \citealt{mauerhan13b}; or even SN~1961V, \citealt{vandyk02,chu04,kochanek11,vandyk12}), and in some cases even the inspection of the sites in deep, high-spatial resolution images obtained many years after the explosion does not provide unequivocal verdicts (e.g. see \citealt{vandyk12}).

The mechanisms triggering these eruptions are still unknown (see \citealt{humphreys94,smith11b}). A connection between some LBVs with SNe IIn as proposed by for example \citep{kotak06,smith06,trundle09}, and occasionally LBVs have been proved to explode as bright SNe IIn \citep[e.g.,][]{galyam07,galyam09}. However, although LBVs are the most usual channel to explain the bursty activity of the SN impostors, these outburst have also been linked to lower mass stars (e.g. the cases of SN~2008S and NGC~3000-OT; \citealt{smith09}, \citealt{bond09}), or the interaction of massive binaries (e.g. \citealt{kashi10}). Outbursts of massive stars may be precursors of terminal SN explosions (e.g. see \citealt{ofek14}), and these instabilities are presumably related to  physical processes occurring when the stars approach the end of their life (for instance, after the beginning of neon or oxygen burning  -- weeks to years prior the explosion; e.g. \citealt{fraser13b,smith14b}). Such outbursts were very likely observed in a few cases, including LSQ13zm \citep{tartaglia16}, SN~2010mc \citep{ofek13}, or the well-studied case of SN~2009ip. The latter had experienced repeated outbursts from - at least - 2009 to early 2012 \citep{pastorello13}, followed by a more luminous, double-peaked re-brightening in summer-autumn 2012 (\citealt{pastorello13,fraser13,mauerhan13a,prieto13,margutti14,graham14}). The mid-to-late 2012 event was interpreted as the terminal core-collapse SN explosion (e.g. \citealt{smith14}), as a merger burst event in a close binary system \citep{soker13,kashi13}, or as collisions of massive shells formed through repeated mass loss events with the progenitor still alive \citep{fraser15,moriya15}. What is clear, in the case of SN~2009ip, is that there is a complex environment surrounding the central object (e.g. \citealt{levesque14,mauerhan14,margutti14,martin15}) which is impeding our inspection of the inner region of the nebula to verify whether the central star is still alive or not. 

In other cases, repeated intermediate-luminosity outbursts have been registered, without leading (so far) to a SN explosion. This sample of rare transients includes SN~2000ch \citep{wagner04,pastorello10}, SNhunt248 \citep{mauerhan15,kankare15}, and UGC 2773-OT \citep{smith16}.

\begin{figure*}
\centering
\includegraphics[width=2\columnwidth]{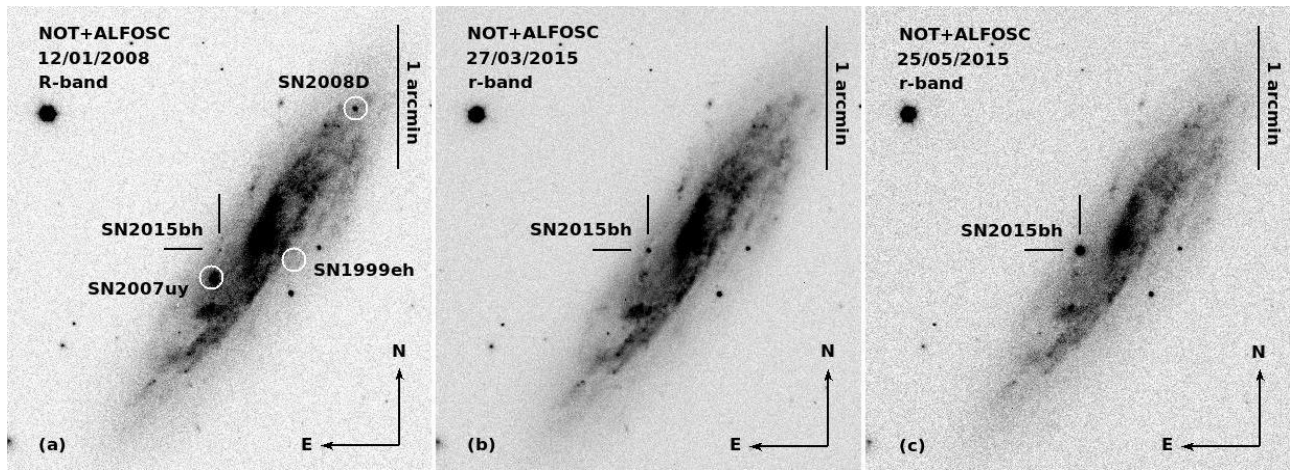}
 \caption{$R$- and $r$-band images of SN~2015bh in NGC~2770 obtained with the 2.56~m Nordic Optical Telescope+ALFOSC at Roque de los Muchachos Observatory (Spain) on 2008 January 12 (a), 2015 March 27 (b), and 2015 May 25 (c). The locations of the transient and those of the three SNe exploded in NGC~2770 are indicated.}
\label{fig_transient}
\end{figure*}

A recent example of a transient with a long variability history is SN~2015bh ($\alpha$=09$^{\rm h}$ 09$^{\rm m}$ 35${\fs}$12, $\delta$=+33$^{\circ}$ 07$\arcmin$ 21${\farcs}$3; J2000.0; Fig. \ref{fig_transient}), also known as SNhunt275, iPTF13efv, PSN J09093496+3307204 or PSN J09093506+3307221. It was discovered in NGC~2770 on 2015 February 07.39 UT, with an unfiltered magnitude of 19.9, by Stan Howerton and the Catalina Real Time Transient Survey (CRTS; \citealt{howerton15})\footnote{\url{http://nesssi.cacr.caltech.edu/catalina/current.html}}, although it was first detected in 2013 by the Intermediate Palomar Transient Factory (iPTF\footnote{\url{http://www.ptf.caltech.edu/iptf/}}; \citealt{ofek16}). Independent discoveries were also reported by Z.-j. Xu (Nanjing, Jiangsu, China) and X. Gao (Urumqi, Xinjiang, China; \citealt{howerton15}). A spectrum was taken on 2015 February 09.93 UTC \citep{eliasrosa15}, by the Asiago Transient Classification Program \citep{tomasella14}. It shows a strong H$\alpha$ emission line with both a broad (FWHM $\sim$ 6800 \kms) and a narrow (FWHM $\sim$ 900 \kms) component, resembling the spectra of the SN/impostors 2000ch \citep{wagner04,pastorello10} and 2009ip (before the explosion of June 2012; e.g. \citealt{pastorello13}).

In this manuscript, we investigate the nature of SN~2015bh. In the next section (section \ref{SNhostgx}), we describe the host galaxy of SN~2015bh. In sections \ref{SNph} and \ref{SNspec}, we present the photometric and spectroscopic results, and in section \ref{SNHST} we constrain the properties of the progenitor star. The combination of all this information is discussed in detail in section \ref{SNnature}. Finally the main results are summarised in section \ref{SNconclus}. Notice that a study on SN~2015bh has already been published by \citet{ofek16}, and further discussed by \citet{soker16} and \citet{thone16}, confirming the complex nature of SN~2015bh. Different possible interpretations of the chain of events of this object are presented in these works (\ref{SNnature}).

%
\section{Host galaxy, distance and reddening of SN~2015bh}\label{SNhostgx}
The host galaxy, NGC~2770, is morphologically classified as a spiral galaxy with a small bulge, open and clumpy spiral arms, and large H\,{\sc i} mass [SA(s)c\footnote{NED, NASA/IPAC Extragalactic Database; \url{http://nedwww.ipac.caltech.edu/}}]. Its star-formation rate is comparable to the values estimated for the Milky Way. The galaxy has a small irregular companion, NGC~2770B, with high star-formation rate (see e.g. \citealt{thone09}). NGC~2770 has already hosted three Type Ib SNe: 1999eh \citep{boles99}, 2007uy, and 2008D (see e.g. \citealt{soderberg08,mazzali08}), being consequently branded as a possible SN Ib factory \citep{thone09}. \\

Throughout the paper, we will adopt a distance to NGC~2770 of $29.3 \pm 2.1$ Mpc ($\mu = 32.33 \pm 0.15$ mag), as derived from the recessional velocity of the galaxy \citep{haynes97} corrected for Local Group infall into the Virgo cluster \citep{mould00} v$_{Vir}$ = $2137 \pm 17$ \kms ($z = 0.007$), and assuming $H_0 = 73$ \kms Mpc$^{-1}$ (values taken from NED).\\

We will also adopt the total reddening value of $E(B-V) = 0.21^{+0.08}_{-0.05}$ mag towards SN~2015bh, as derived by \citet{thone16} from the equivalent width of the interstellar Na ID lines in a high-resolution spectrum taken on 2015 June 04.

%

\section{Photometry}\label{SNph}

\subsection{Observations and data reduction}\label{SNph_red}

Optical {\it UBVRI} (Johnson Cousins system) and {\it ugriz} (Sloan system) images of SN~2015bh were taken using a large number of observing facilities, listed in Table \ref{table_setup}. We also collected archival and unfiltered data from amateur astronomers taken since 1994, i.e. $\sim$ 21 yr before the target discovery. The data set was completed with images taken in the near infrared (NIR) domain, and observations from space telescopes such as the Ultraviolet and Optical Telescope (UVOT) on board of the {\sl SWIFT} satellite, and the {\sl Hubble Space Telescope\/} ({\sl HST}).

Photometric observations were pre-processed following the standard recipe in {\sc iraf}\footnote{{\sc iraf} is distributed by the National Optical Astronomy Observatory, which is operated by the Associated Universities for Research in Astronomy, Inc., under cooperative agreement with the National Science Foundation.} for CCD images (overscan, bias, and flat-field corrections). For infrared exposures, we also applied an illumination correction and sky background subtraction using the external {\sc iraf} package NOTCam (version 2.5)\footnote{\url{http://www.not.iac.es/instruments/notcam/guide/observe.html}} for the NOT images and a custom IDL routines for the CPAPIR images \citep{artigau04}. The SN magnitudes were measured using a dedicated pipeline \citep[SNOoPY;][]{cappellaro14}. This consists of a collection of {\sc python} scripts calling standard {\sc iraf} tasks (through {\sc pyraf}), and other specific analysis tools, in particular {\sc sextractor}, for instrument extraction and star/galaxy separation, {\sc daophot}, to measure the instrumental magnitude via point-spread function (PSF) fitting, and {\sc hotpants}\footnote{\url{http://www.astro.washington.edu/users/becker/hotpants.html}}, for image difference with PSF matching.

In order to calibrate the transient's instrumental magnitudes to standard photometric systems, we used SDSS stars in the field as reference. When needed, these were converted to the Johnson Cousins system using the relations in \citet{chonis08}. For the infrared photometry, we used as reference for the calibration the Two Micron All Sky Survey (2MASS) catalog.

Unfiltered instrumental magnitudes from amateur images were also measured through the PSF fitting technique. These were then rescaled to Sloan {\it r}-band magnitudes, as this best matches the quantum efficiency peaks of the detectors used for these observations. 

When the transient was not detected, upper limits were estimated, corresponding to a peak of 2.5 times the background standard deviation. Error estimates were obtained through an artificial star experiment, combined (in quadrature) with the PSF fit error returned by {\sc daophot}, and the propagated errors from the photometric calibration.

The final calibrated magnitudes of SN~2015bh are listed in Tables \ref{table_JCph}, \ref{table_SLph}, and \ref{table_NIRph}. Optical {\it UBVRI} and NIR data are reported in {\sc vegamag} scale, while {\it ugriz} data are in {\sc ab mag} scale.

{\sl SWIFT} pointed to the field of SN~2015bh at different epochs since 2008 with ultraviolet (UV) and optical filters thanks to the follow-up campaigns of SNe~2007uy and 2008D. A preliminary analysis showed that the transient was not visible in 2008. We estimated upper limits of 19.6, 19.6, and 19.4 mag for {\sl SWIFT} $UVW2$, $UVM2$, and $UVW1$ respectively. We therefore combined all images of 2008, and used the resulting stacked image as a template for the analysis of images obtained in subsequent epochs. The magnitudes of the transient were obtained using the pipeline from the Swift's Optical/Ultraviolet Supernova Archive ({\sc SOUSA}; \citealt{brown14}), which uses revised zero-points on the UVOT-Vega system \citep{breeveld11} and includes time-dependent sensitivity corrections. The derived magnitudes are listed in Table \ref{table_UVph}. As the UVOT {\it U} band is much bluer than Johnson Cousins {\it U} or Sloan {\it u}, we will treat these bands separately. 

Finally, {\sl HST} observed the SN~2015bh field with WFPC2 in a large set of filters between 2008 and 2009 (see Table \ref{table_HSTph}). The magnitudes of the transient in \textsc{vegamag} were obtained using the {\sc HSTphot}\footnote{{\sc HSTphot} is a stellar photometry package specifically designed for use with {\sl HST\/} WFPC2 images. We used v1.1.7b, updated 8 September 2009. \url{http://americano.dolphinsim.com/hstphot/}} package \citep{dolphin00}.

\subsection{Light curves}\label{SNph_lc}

\begin{figure*}
\centering
\includegraphics[width=1.3\columnwidth]{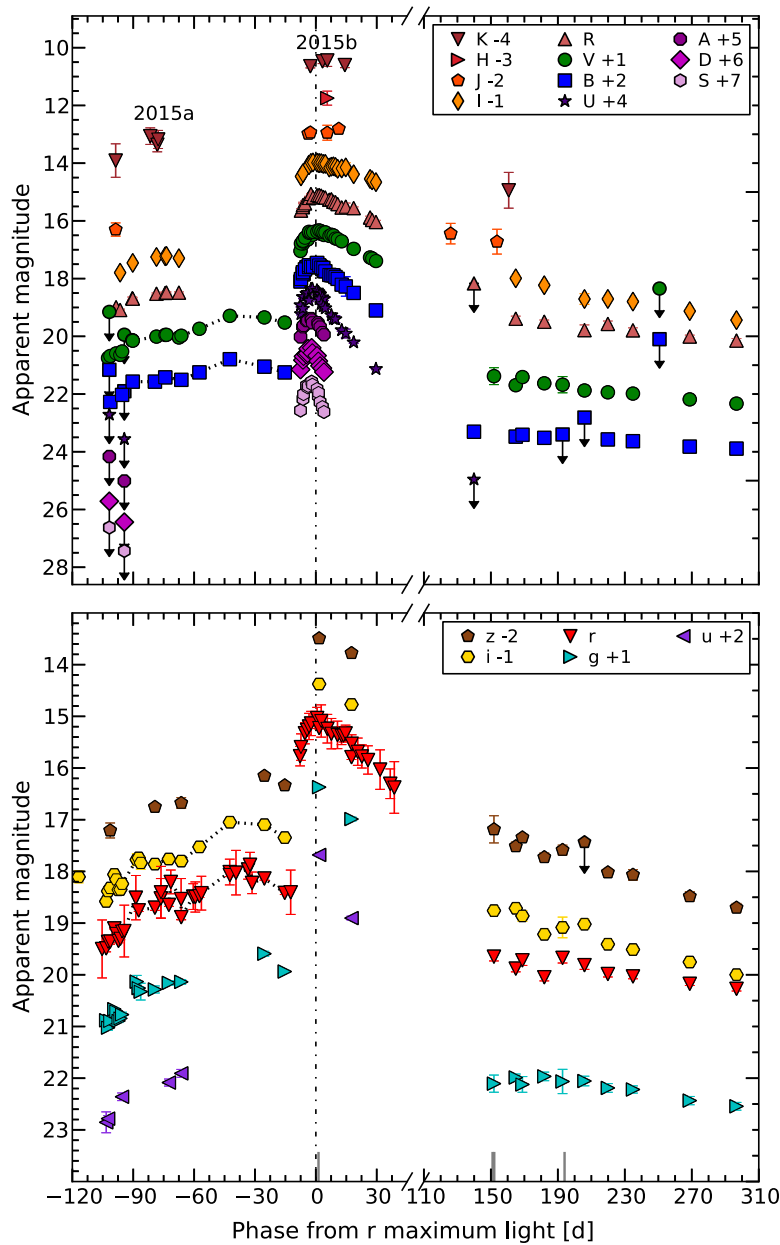}
  \caption{Optical light curves of SN~2015bh. Upper limits are indicated by a symbol with an arrow. The solid marks on the abscissa axis indicate the phases at which spectra are obtained. The dotted line connects the magnitudes during the 2015a event. The dot-dashed vertical line indicates the $r$-band maximum light of SN~2015bh. The light curves have been shifted for clarity by the amounts indicated in the legend. Note that the filters named as $SDA$ correspond to the {\sl SWIFT} $UVW2,UVM2,UVW1$ filters, respectively. The uncertainties for most data points are smaller than the plotted symbols. A colour version of this figure can be found in the online journal.}
\label{fig_ph}
\end{figure*}

The UVOT $UV$ and $uUBgVrRiIzJHK$ light curves of SN~2015bh after the discovery on 2015 February 07 are shown in Fig.~\ref{fig_ph}. The light curve of the transient shows a slow rise of $\sim$ 1.5 mag in around 100 d in all bands, which we label as the `2015a' event. This episode is followed by a sudden steeper increase in the light curve brightness (by about 3 magnitudes in less than 10 days), labelled as the `2015b' event (this re-brightening was also reported by \citealt{deugarte15}, and independently detected by R. Arbour\footnote{\url{http://www.cbat.eps.harvard.edu/unconf/followups/J09093496+3307204.html}}, South Wonston, U.K.). This leads to a light curve peak of $-$17.81 mag in the $r$ band, followed by a slow decline for the next 30 days. The observational campaign was subsequently interrupted because of the alignment with the Sun, and observations restarted about 3 months later. At that time, the transient was still visible but had dimmed by over 4 magnitudes (see Table \ref{table_JCph}). We also notice that, after the 2015b peak, the luminosity of the object decreases more rapidly in the blue bands than the red bands, indicating that the peak of the spectral energy distribution (SED) progressively shifts to longer wavelengths. Table \ref{table_max} reports the peak magnitudes of the 2015a event, as well as the peak epochs and magnitudes for the 2015b event, all obtained by fitting the light curves with low-order polynomials. Post-maximum and tail decline rates are also disclosed in the same table. In the following, we will adopt as reference epoch that of the 2015b $r$-band maximum, i.e. 2015 May 24.28, or MJD $57166.28 \pm 0.29$. 

\begin{figure*}
\centering
\includegraphics[width=1.8\columnwidth]{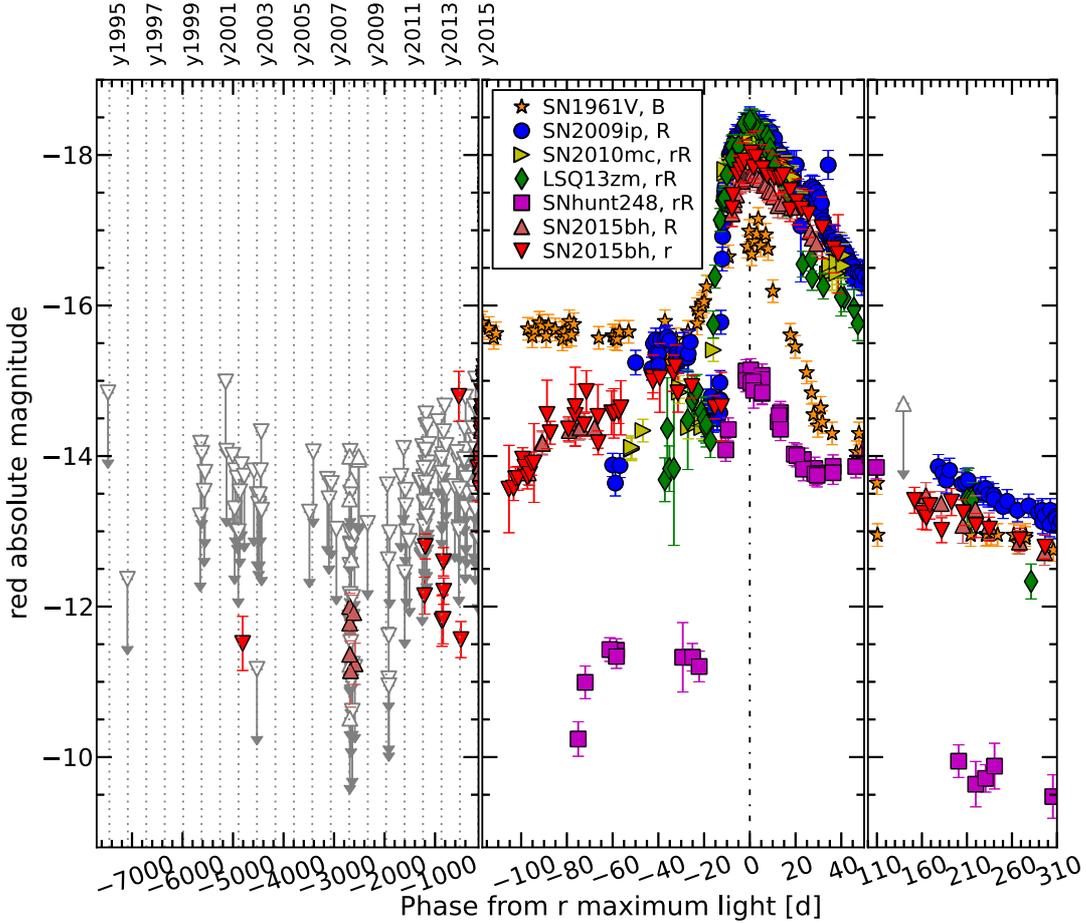}
 \caption{Historical absolute {\sc vegamag} $rR$ light curve of SN~2015bh ({\it filled triangles}), shown along with those of  SNe~1961V (in $B$-band; {\it stars}), 2009ip ({\it circles}), 2010mc ({\it rotated triangles}), LSQ13zm ({\it diamonds}), and SNhunt248 ({\it squares}). SN~2015bh's upper limits are indicated by empty triangles with arrows. The dot-dashed vertical line indicates the $r$-band maximum light of SN~2015bh. A colour version of this figure can be found in the online journal.}
\label{fig_abs}
\end{figure*}

As we mentioned before, the site of SN~2015bh was monitored for more than 20 yr before the transient's discovery (see Fig. \ref{fig_transient} and tables \ref{table_JCph}, \ref{table_SLph}, \ref{table_UVph} and \ref{table_HSTph}). A large fraction of data was collected by amateur telescopes, complemented by a few deep images obtained with the Isaac Newton Telescope (see Table \ref{table_setup}) and the Pan-STARRS telescope \citep{kaiser10} during its 3$\Pi$ survey operations \citep[the filter system and calibration are described in][]{magnier13,schlafly12,tonry12}. In these deeper images, we detect a source at the position of SN~2015bh. In addition, examining the data from the extensive follow-up campaigns of SNe~2007uy and 2008D, additional detections are found in 2008 and 2009, including data taken with {\sl HST}. Interestingly, the transient is detected only at red wavelengths during these years from ground-based telescopes, and we could obtain only upper limits in the other bands. In Fig. \ref{fig_abs} we plot the historic $rR$ absolute light curve of SN~2015bh ($r$ magnitudes in the Sloan system have been scaled to the {\sc vegamag} system by adding a conversion value of 0.16; \citealt{blanton07}). The pre-discovery detections and upper limits indicate that the target likely remained at a magnitude below $-$14 for almost 21 yr, except for an outburst at $r \sim -14.5$ mag in December 2013 (detected with a 0.4m telescope), coincident with the iPTF detection \citep{duggan15,ofek16}. Neglecting this outburst, we may note a long-duration brightening, and some signatures of erratic variability.

The $rR$ absolute magnitude light curve of SN~2015bh is compared in Fig. \ref{fig_abs} with those of other objects with multi-peaked light curves, namely the controversial SNe~1961V \citep{bertola63,bertola64,bertola65,bertola67}, 2009ip \citep{maza09,pastorello10,smith10,foley11,pastorello13,mauerhan13a,fraser13,margutti14,fraser15}, SN~2010mc \citep{ofek13}, LSQ13zm \citep{tartaglia16}, and SNhunt248 \citep{kankare15}. Only the latter was clearly an impostor, although its multi-peaked light curve shows some resemblance with the other objects of the sample. The light curves of the comparison objects have been computed accounting for the distance and extinction values obtained from the literature\footnote{SN~1961V: ${E(B-V)}_{\rm tot} = 0.05$ mag, $\mu = 29.84$ mag; SN~2009ip: ${E(B-V)}_{\rm tot} = 0.02$ mag, $\mu = 32.05$ mag; SN~2010mc: ${E(B-V)}_{\rm tot} = 0.01$ mag, $\mu = 35.79$ mag; LSQ13zm: ${E(B-V)}_{\rm tot} = 0.02$ mag, $\mu = 35.43$ mag; SNhunt248: ${E(B-V)}_{\rm tot} = 0.05$ mag, $\mu = 31.76$ mag.}. From the comparison, we note that both the 2015a and 2015b events of SN~2015bh are fainter than the equivalent events observed in SNe 2009ip, 2010mc and LSQ13zm, but brighter than those of the impostor SNhunt248. All the precursor outbursts of these transients show shorter duration than 2015a, except for SN~1961V. In general, the absolute magnitude of SN~2015bh during the 2015b event falls in the interval of peak magnitudes observed in SNe~IIn, which is between $-16$ and $-19$ mag (\citealt{kiewe12}; the 2015a event reaches a maximum value of $\sim -14.8$ mag in the $r$ band). It is significantly brighter than SNhunt248, which is one of the most luminous confirmed SN impostors. All of this may suggest that the 2015b event was an actual SN explosion. However, as we will see in section \ref{SNspec}, spectroscopic considerations may lead to different conclusions.

One additional property is that, during the 2015a event, the SN~2015bh light curves possibly show some small-scale fluctuations superposed on the broad curvature (see Fig. \ref{fig_ph}). A more evident modulation was seen in the SN~2009ip light curve, although after the 2012b peak (e.g. \citealt{martin15}), and was attributed to clumps or heterogeneity in the gas shells expelled by the progenitor star in previous mass-loss events\footnote{\citet{soker13} and \citet{kashi13} suggest that these fluctuations are consequence of the interaction between shells of material excreted from a progenitor binary system during periastron passage. }

Finally, around 150 d after maximum, SN~2015bh has faded to $M_r = -13.26\pm 0.17$ mag, $\sim$4 mag dimmer than at maximum, showing a slow decline, mostly notable in the redder bands (see Table \ref{table_max}). This behaviour suggests a still ongoing CSM interaction. The luminosity at these phases is $\sim$2 mag higher than at the first detection of SN~2015bh in March 2002 (see Table \ref{table_SLph}). Moreover, as we can see in Fig. \ref{fig_abs}, SN~2015bh remains always fainter than SN~2009ip at coeval epochs.\\

\begin{figure}
\centering
\includegraphics[width=1\columnwidth]{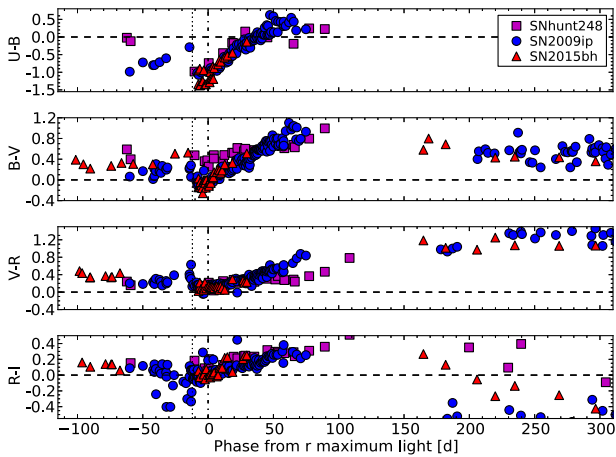}
\caption{Intrinsic colour curves of SN~2015bh ({\it filled triangles}), compared with those of  SN~2009ip ({\it circles}), and SNhunt248 ({\it squares}). The dotted vertical line marks the approximate date of the beginning of the 2015b event. The dot-dashed vertical line indicates the $r$-band maximum light of SN~2015bh. A colour version of this figure can be found in the online journal. }
\label{fig_color}
\end{figure}

In Fig. \ref{fig_color}, we show the evolution of intrinsic Johnson-Cousins colour indices for SN~2015bh, SN~2009ip, LSQ13zm and SNhunt248. SN~2015bh shows a relatively flat colour evolution during the 2015a event, but suddenly becomes bluer when the 2015b event takes place. After maximum it turns again toward the red and at late times, i.e. $\gtrsim$ 150 d after the maximum light, the $(B-V)_0$ colour of SN~2015bh is roughly the same as that registered during the 2015a event. A similar colour evolution is also seen in the comparison objects during the most luminous outbursts, whilst there are some differences during the first event. \\

\begin{figure}
\centering
\includegraphics[width=1.05\columnwidth]{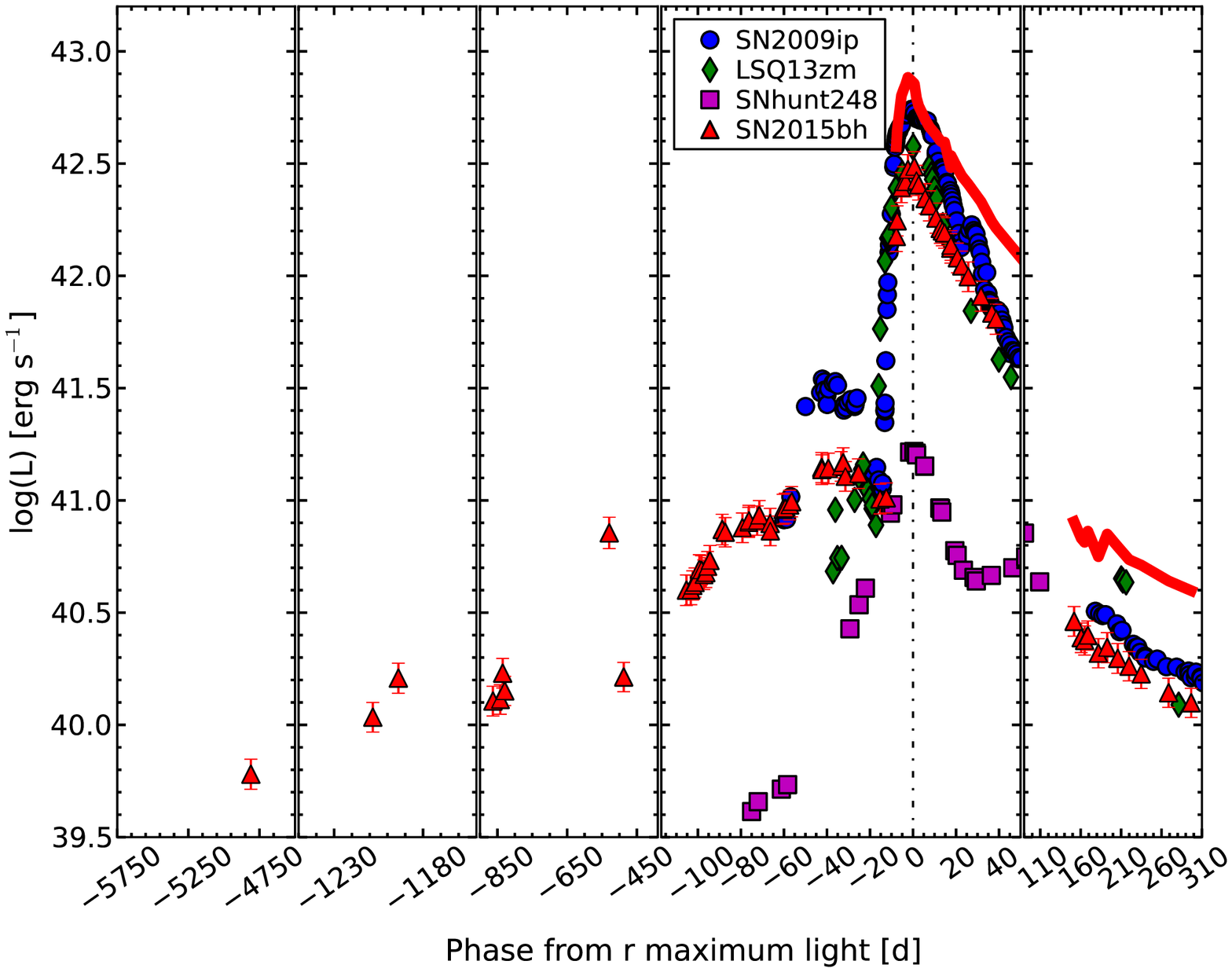}
\caption{Pseudo-bolometric optical light curves of SN~2015bh ({\it triangles}) compared with those of SN~2009ip ({\it circles}), LSQ13zm ({\it diamonds}), and SNhunt248 ({\it squares}). The UV-optical-NIR light curve of SN~2015bh during the 2015b event is also shown ({\it thick solid line}). The dot-dashed vertical line indicates the $r$-band maximum light of SN~2015bh. A colour version of this figure can be found in the online journal.}
\label{fig_bol}
\end{figure}

We computed a pseudo-bolometric light curve of SN~2015bh for each event separately (see Fig. \ref{fig_bol}). The fluxes at the effective wavelengths were derived from extinction-corrected apparent magnitudes. We computed the bolometric luminosity at epochs when observations in the $r$ band were available. When no observation in another filter was available, the missing photometric point was recovered by interpolating the values from epochs close in time or, when necessary, by extrapolating the missing photometry assuming a constant colour. We estimated the pseudo-bolometric flux at each epoch integrating the SED using the trapezoidal rule, and assuming zero flux outside the integration boundaries. Finally, the luminosity was derived from the measured flux accounting for the adopted distance. For phases $< -10$ d, we integrated the flux only for the optical wavelength range, i.e., from $U$ to $z$ band, while for the 2015b event, we computed the pseudo-bolometric curve considering first the optical bands only, and then including the $UV$ and NIR bands.

The errors in the bolometric luminosity account for the uncertainties in the distance estimate, the extinction and the apparent magnitudes. 
 
By fitting low order polynomials to the pseudo-bolometric light curve, we estimated that SN~2015bh reached a peak of luminosity of (1.4 $\pm$ 0.3) $\times$ 10$^{41}$ erg s$^{-1}$ during the first event, and (29.3 $\pm$ 6.1) $\times$ 10$^{41}$ erg s$^{-1}$ during the 2015b event -- which increases to (72.9 $\pm$ 16.0) $\times$ 10$^{41}$ erg s$^{-1}$ if we include the $UV$ and the NIR contribution (see Table \ref{table_lum}). In Fig. \ref{fig_bol}, we include also the pseudo-bolometric light curves (from $U$ to $z$ band) of SN~2009ip, LSQ13zm and SNhunt248, which we calculated in a similar manner as that of SN~2015bh (for LSQ13zm we have no contribution estimated in the $U$ band). As we can see in the figure, while the overall luminosity of SN~2015bh is fainter than those of SN~2009ip and LSQ13zm (but more luminous than SNhunt248). 
 
\begin{table*}
 \centering
  \caption{Peak epochs, peak apparent magnitudes, and decline rates of SN~2015bh in different bands.}
  \label{table_max}
  {
  \begin{tabular}{@{}lcccccc@{}}
     \hline
     Band$^a$ & MJD$_\mathrm{max,2015a}$ & m$_\mathrm{max,2015a}$ & MJD$_\mathrm{max,2015b}$ & m$_\mathrm{max,2015b}$ & Decline from max. & Tail rate at $\gtrsim$ 150 d\\
     &  & (mag) &  & (mag) & [mag (30d)$^{-1}$]$^{b}$ & [mag (100d)$^{-1}$] \\
     \hline
     $U$ & - & - &  57165.04 (0.10)  & 14.67 (0.05) & 2.81 (0.05) & - \\ 
     $B$ &  57131.35 (2.03) & 18.87 (0.06) & 57165.34 (0.33) & 15.50 (0.05)  & 1.27 (0.05) & 0.31 (0.05)\\
     $V$ &  57134.07 (1.00) & 18.28 (0.06) & 57167.03 (0.11) & 15.38 (0.05)  & 1.07 (0.05) & 0.72 (0.13)\\
     $R$ & - &  - & 57166.71 (0.20) & 15.11 (0.05) & 0.95 (0.05)  & 0.52 (0.21)\\
     $I$ & - & - & 57166.26 (0.13) & 14.98 (0.05) & 0.70 (0.05)  &  1.21 (0.20)\\
     \hline
     $u$ & - &  - & - & -  & 2.28 (0.50) & - \\
     $g$ & - &  - & - & -  & 1.17 (0.25)  & 0.38 (0.07)\\
     $r$ &  57132.35 (1.00) & 18.11 (0.06) & 57166.28 (0.29) & 15.10 (0.20)  & 1.06 (0.16) & 0.46 (0.07)\\
     $i$ &  57134.35 (0.59) & 18.07 (0.06) & - & -  & 0.73 (0.20) & 1.01 (0.09)\\
     $z$ & 57136.00 (1.00) & 18.19 (0.06) & - & -  & 0.55 (0.20) & 1.03 (0.16)\\
     \hline
     $J$ & - &  - & 57170.69 (1.20) & 14.81 (0.20) & - & - \\
     $K$ & - &  - & 57172.87 (0.62) & 14.44 (0.20) & - & - \\
     \hline
    $UVW2$ & - & - & 57163.92 (0.23)  & 14.62 (0.05) & 7.02 (0.11) & -\\ 
    $UVM2$  & - & - & 57163.56 (0.31)  & 14.48 (0.05) & 5.00 (0.06) & -\\
    $UVW1$  & - & - & 57163.46 (0.35)   & 14.42 (0.05) & 4.10 (0.06) & -\\
    \hline
  \end{tabular}
  }
  \begin{flushleft} 
  $^a$ The maximum magnitude of the 2015a event of the $uUgRIJHK$ and ultraviolet light curves could not be constrained. The same is true for the $ugizH$ peaks of the 2015b event.\\
  $^b$ Considering an interval of 30 d from maximum light. In the case of $ugiz$, we extrapolate to 30 d the decline estimated between the only 2 detections in these bands after maximum.\\
  \end{flushleft}
\end{table*}

\begin{table}
 \centering
  \caption{Peak of the pseudo-bolometric$^a$ light curves of SN~2015bh and comparison transients.}
  \label{table_lum}
  {
 \begin{tabular}{@{}lcc@{}}
\hline
 Object & Luminosity$_\mathrm{max,2015a}$ & Luminosity$_\mathrm{max,2015b}$ \\
 & ($\times\ 10^{41}$ erg s$^{-1}$) & ($\times\ 10^{41}$ erg s$^{-1}$) \\
\hline
SN~2009ip & - & 53.0 (7.3) \\
LSQ13zm & - & 38.6 (5.4) \\
SNhunt248 & - & 1.6 (0.2) \\
SN~2015bh & 1.4 (0.3) & 29.3 (6.1) \\
\hline
SN~2015bh$^b$ & - & 72.9 (16.0) \\
\hline 
\end{tabular}}
\begin{flushleft}
$^a$  From $U$ to $z$ bands but for LSQ13zm in which the $U$ band was not available.\\
$^b$ Including the $UV$ and the NIR contribution. \\
\end{flushleft}
\end{table}

%

\section{Spectroscopy}\label{SNspec}

\subsection{Observations and data reduction}\label{SNspec_red}

Spectroscopic monitoring of SN~2015bh started soon after the discovery, on 2015 February 09.97 UT, and lasted until 2016 March 05.89 UT, interrupted by $\sim$ 100 days when the transient was too close to the Sun. Basic information on our spectra is reported in Table \ref{table_spec}. 

All spectra were reduced following standard procedures with {\sc iraf} routines. The two-dimensional frames were de-biased and flat-field corrected, before performing the extraction of the 1D spectra. The wavelength calibration was accomplished with the help of arc-lamp exposures obtained in the same night, and then the accuracy of the calibration was checked using night-sky lines. The spectra were flux calibrated using the high signal-to-noise exposure of spectrophotometric standards stars \citep{oke90,hamuy92,hamuy94}. Finally, the flux calibrated spectra were checked against the photometry at coeval epochs and a correction factor was applied to the flux in case of discrepancy. The standard star spectra were also used to remove the strongest telluric absorption bands  (in some cases, residuals are still present after the correction). 

\subsection{Evolution of the spectral continuum and the major features}\label{SNspec_evol}

\begin{figure*}
\centering
\includegraphics[width=1.9\columnwidth]{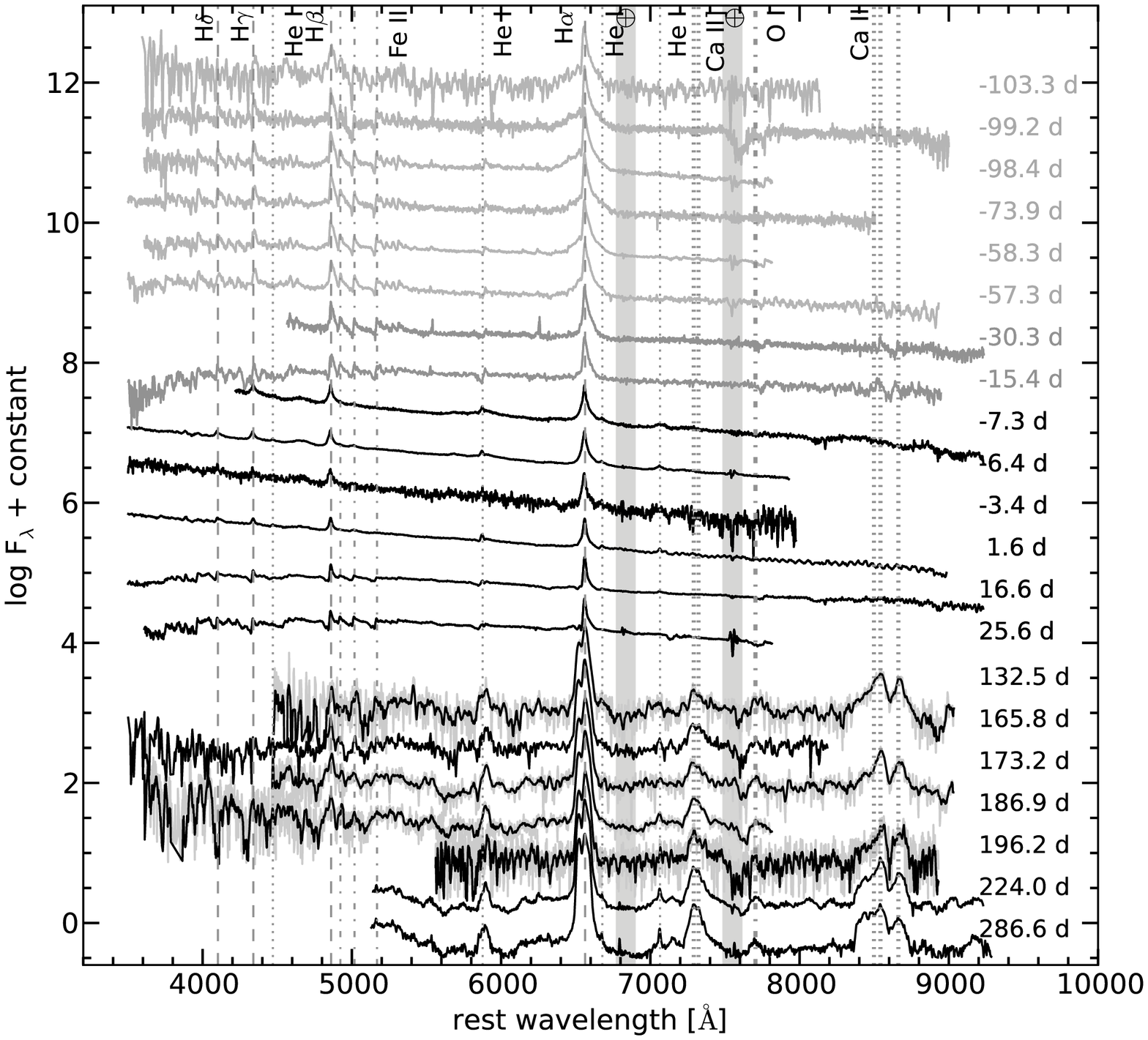}
 \caption{Sequence of optical spectra of SN~2015bh taken from 2015 February 09.97 UTC to 2016 March 05.89 UT. Shades of grey are used for spectra obtained during the 2015a event, spectra taken during and after the 2015b event are in black. The late spectra at 132.5, 173.2, 186.9, and 196.2 d are shown in grey, with a boxcar-smoothed (using a 8 pixel window) version of the spectra overplotted in black. The locations of the most prominent spectral features are indicated by vertical lines.}
\label{fig_optspec}
\end{figure*}

Fig. \ref{fig_optspec} shows the sequence of optical spectra of SN~2015bh. 

During the early stages of the 2015a event, i.e. from $-100$ d to $-60$ d, the spectra exhibit a mildly blue continuum and very little evolution. They are dominated by multi-component Balmer lines in emission and Fe\,{\sc ii} features. Instead, the spectrum at $-15.4$ d (our last spectrum of the 2015a event), shows a red continuum, indicating that the temperature of the emitting regions has decreased (more detail is given below). However, at the onset of the 2015b event, the continuum temperature changes drastically, increasing by a factor two, and then cools down again when the luminosity declines. During the first days after maximum light, the only visible features in the spectra are the Balmer lines, along with weak He\,{\sc i} $\lambda$5876 (possibly blended with Na\,{\sc i}), $\lambda$6678, and $\lambda$7065 features. From +16 d onwards, when the continuum becomes redder, the He\,{\sc i} lines fade in intensity and the Fe\,{\sc ii} line forest reappears. In particular, we note at some early epochs ($< 30$ d) the presence of one or two weak absorptions features on the blue side of the H$\alpha$ line.

At late phases (>130 d), when the transient was recovered after the seasonal gap, the spectra show a deep change. The narrow lines have disappeared and broad lines are now evident. The He\,{\sc i} features are more intense, and lines of calcium such as Ca\,{\sc ii}] $\lambda\lambda$7291, 7323, and Ca\,{\sc ii} $\lambda\lambda\lambda$8498, 8542, 8662, along with sodium (Na\,{\sc i} $\lambda\lambda$5891, 5897; possible blended with He\,{\sc i} $\lambda$5876) and weak oxygen (O\,{\sc i} $\lambda$7774 and $\lambda$8446, and [O\,{\sc i}] $\lambda$5577 and $\lambda\lambda$6300, 6364) are also present. The H$\alpha$ profile has also changed showing now three components. See section \ref{latetime} for more details.
\\

\begin{figure}
\centering
\includegraphics[width=1.\columnwidth]{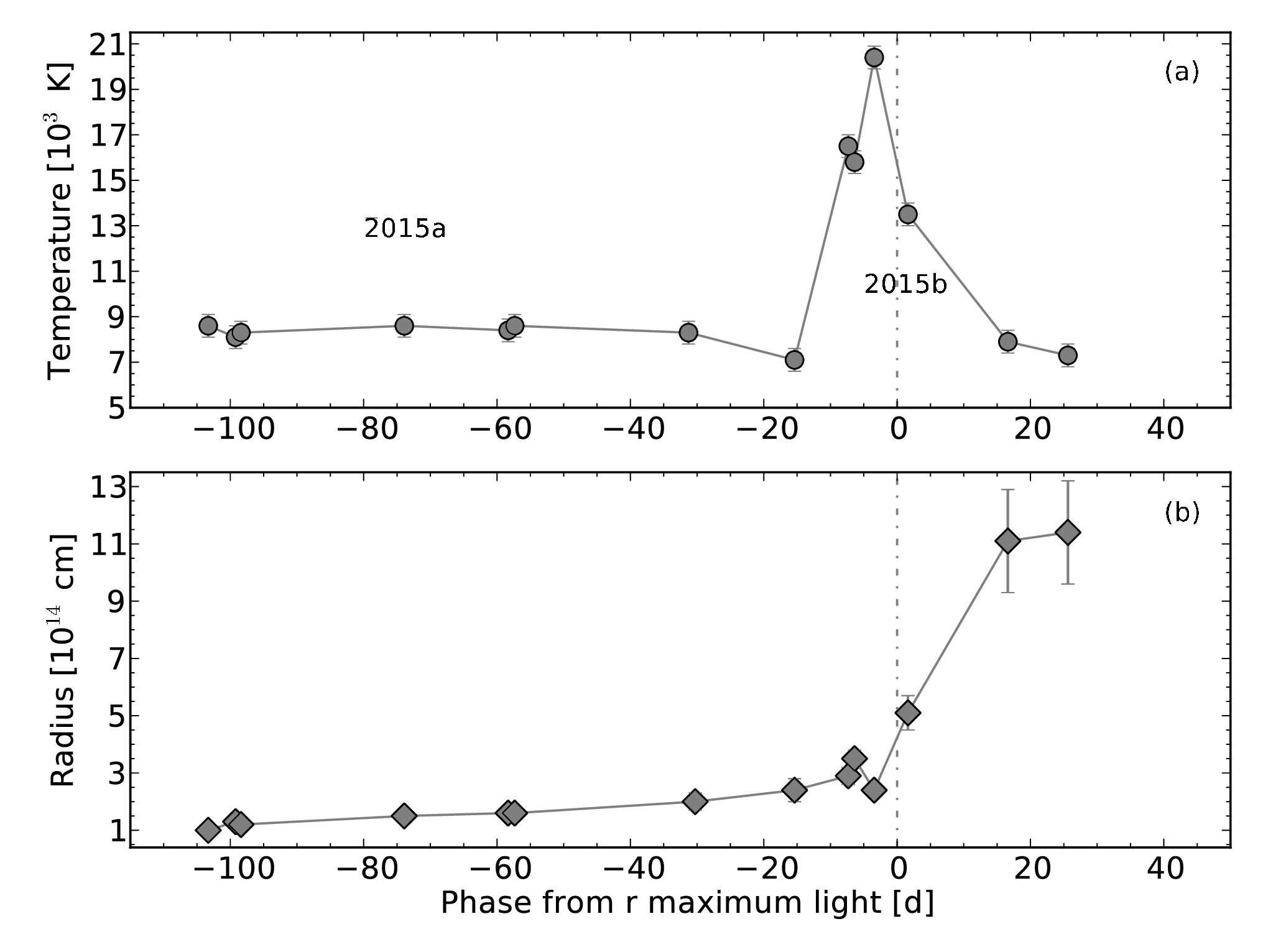}
 \caption{Panel (a): Evolution of the best-fit black-body temperatures. Panel (b): Evolution of the radius at the photosphere. The solid line connects the temperature and radius measurements. The dot-dashed vertical line indicates the $r$-band maximum light of SN~2015bh (MJD$_{max}$ = $57166.28 \pm 0.29$). Both 2015a and 2015b events are also indicated.}
\label{fig_spectemp}
\end{figure}

The photospheric temperature is estimated by fitting the SED of SN~2015bh with a black-body function after removing the strongest features of the spectra. The temperature evolution of SN~2015bh is shown in the panel (a) of Fig. \ref{fig_spectemp}. A conservative uncertainty for the temperature of about $\pm$ 500 K is assumed in our temperature estimates. As aforementioned, it increases from an average $T_\mathrm{bb}$ of 8500 K during the 2015a event, to $T_\mathrm{bb}$ $\sim$ 20000 K at the peak of the 2015b event, after passing through a short-lasting temperature minimum, at $T_\mathrm{bb}$ $\sim$ 7100 K, just before the major re-brightening (see also Fig. \ref{fig_optspec}). Within approximately 15 d from maximum light, the temperature decreases again to a similar value as during the 2015a event, i.e., $\sim$ 8000 K. These values are comparable with those of SN~2009ip (e.g. \citealt{margutti14}), or luminous interacting SNe IIn \citep{taddia13}, but somewhat higher (by around 30 per cent) than those of LSQ13zm \citep{tartaglia16}. 
\\

Given the temperature and the pseudo-bolometric luminosity of SN~2015bh, we approximat the evolution of the radius of the photosphere. As displayed in panel (b) of Fig. \ref{fig_spectemp}, the photospheric radius could have a very slow increase from 1 $\times$ 10$^{14}$ to 3 $\times$ 10$^{14}$ cm in about 100 d. Then, the radius sharply increases during the re-brightening of SN~2015bh (2015b). The same overall behaviour of the radius is also observed, for instance, in SNe~2009ip. 

Note that this is a rough estimation of the photospheric radius since we are making assumptions in deriving the temperature and the luminosities of SN~2015bh. For instance, we are assuming a black-body spectrum where the real spectra are also affected by the metal line blending, while we integrate the luminosity over a limited range of wavelength (from $U$ to $z$ band).

%
\subsection{Evolution of the Balmer lines}\label{SNspec_evolhalpha}

\begin{figure}
\centering
\includegraphics[width=0.75\columnwidth]{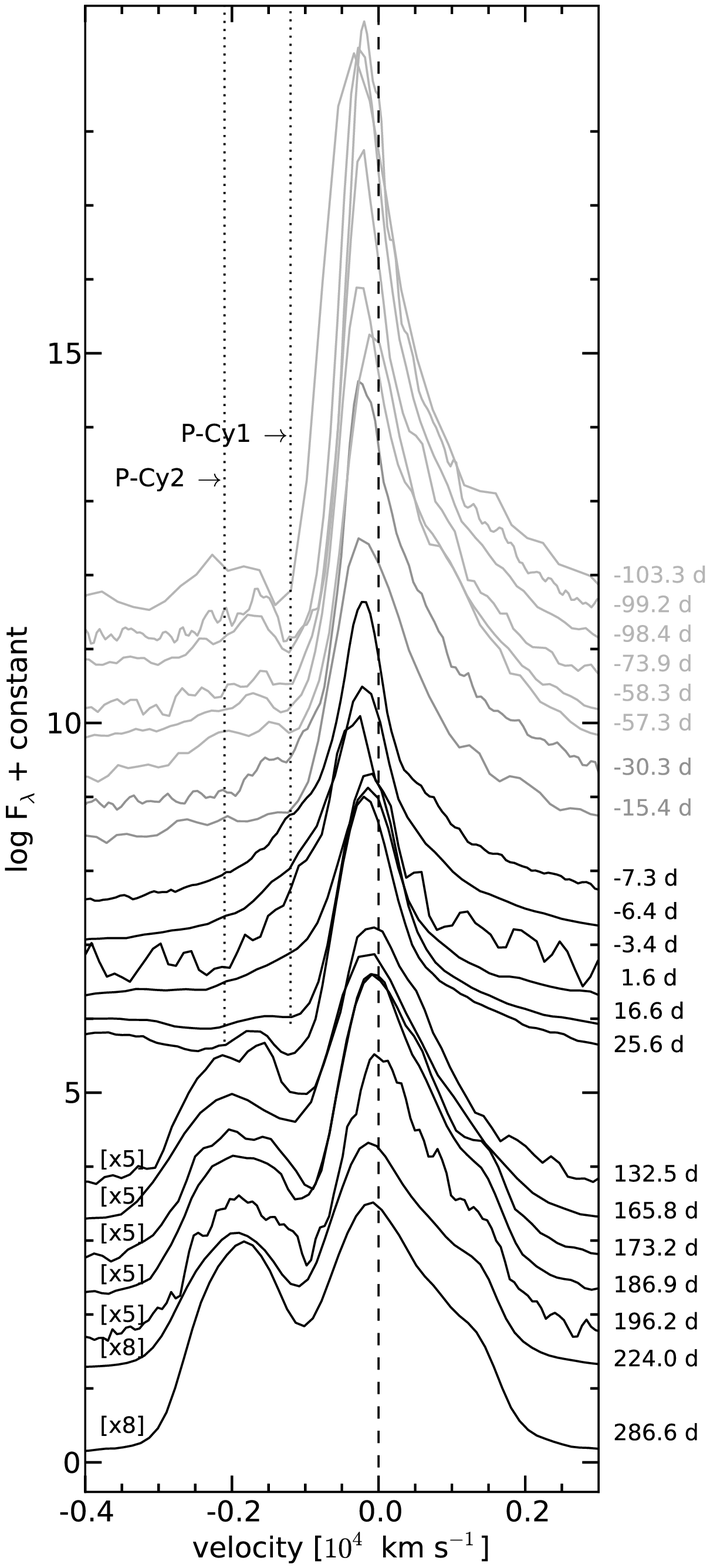}
 \caption{Evolution of the H${\alpha}$ profile in velocity space. The dotted lines indicate the major blue absorption components distinguished in the H${\alpha}$ line profile. The dashed lines mark the rest wavelength of H${\alpha}$. Spectra have been vertically shifted for clarity by an arbitrary amount. 
 }
\label{fig_spechalpha}
\end{figure}

\begin{figure}
\centering
\includegraphics[width=1.\columnwidth]{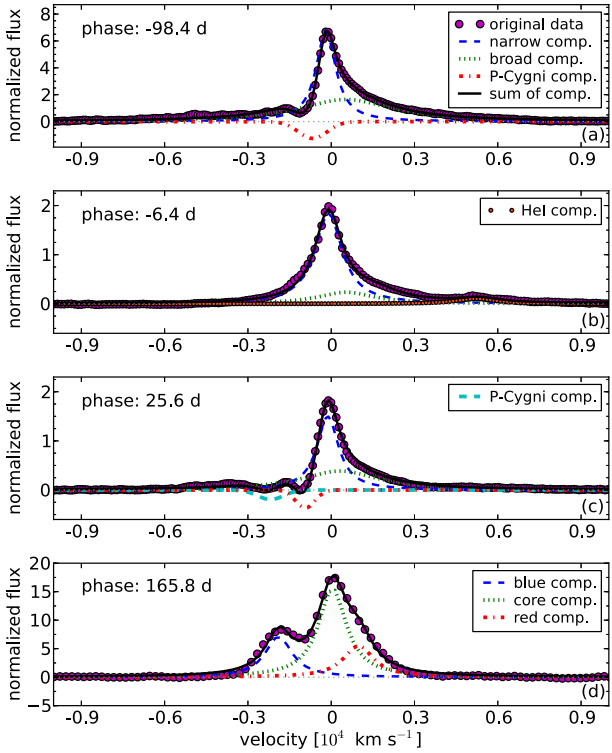}
\caption{Decomposition of the H${\alpha}$ emission line of SN~2015bh before ($-98.4$ d), during ($-6.4$ d) and after (26.1 d and 165.8 d) the 2015b event. A colour version of this figure can be found in the online journal.}
\label{fig_specdecomp}
\end{figure}

\begin{figure}
\centering
\includegraphics[width=1.\columnwidth]{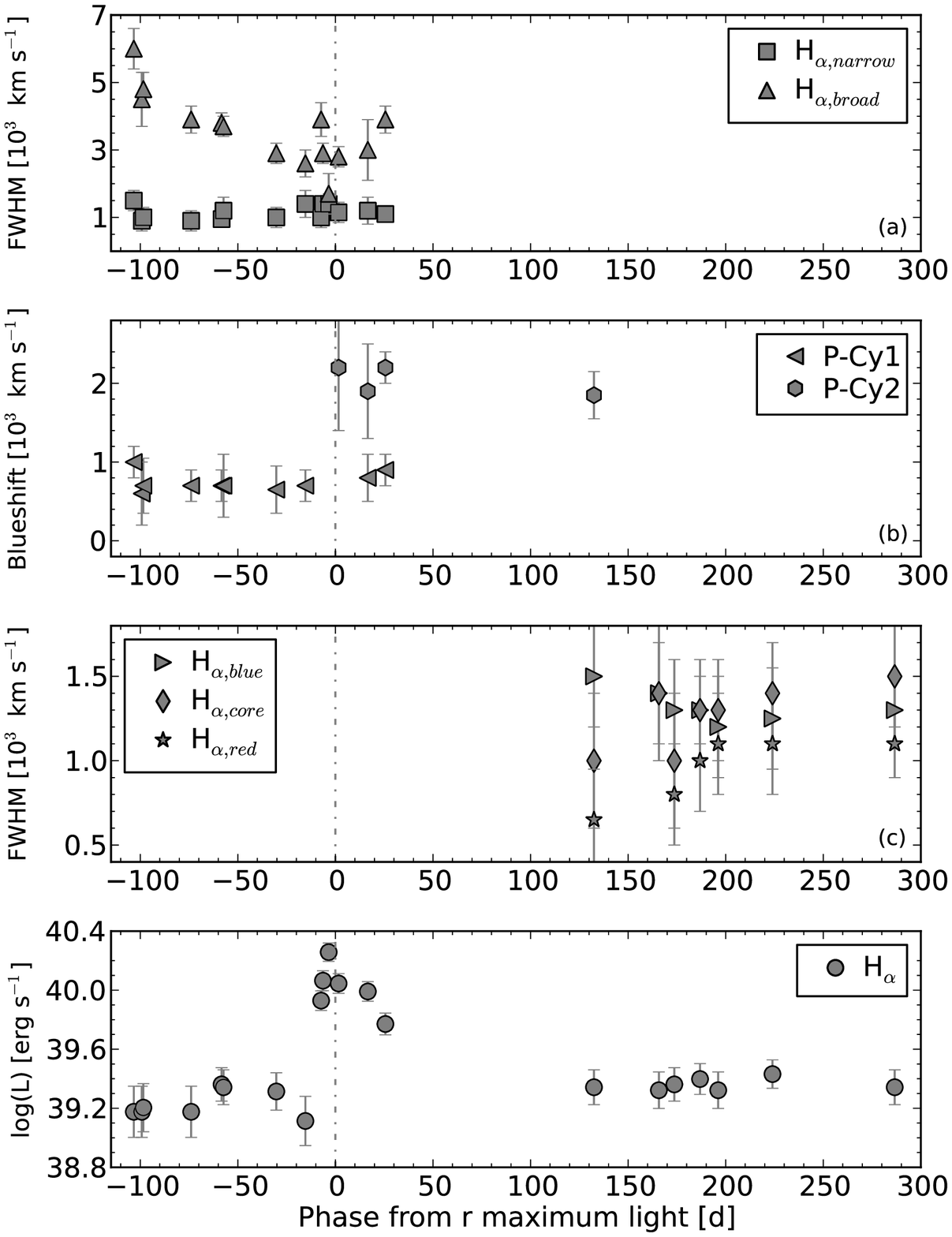}
 \caption{Panel (a): FWHM evolution for the broad and narrow H${\alpha}$ emissions. Panel (b): Evolution of the blueshift of the P-Cy1 and P-Cy2 absorptions. Panel (c): FWHM evolution for the blue, core, and red H${\alpha}$ components at late phases. Panel (d): Evolution of the total luminosity of H${\alpha}$. The dot-dashed vertical line indicates the $r$-band maximum of SN~2015bh.}
\label{fig_specvel}
\end{figure}

The Balmer line profiles, in particular those of H$\alpha$, show evident changes in morphology with time (Fig. \ref{fig_spechalpha}). In particular, we note strong differences in the line profiles between early and late phases. Analysing the evolution of the H$\alpha$ components may allow us to probe the transient's environment, and hence the nature of SN~2015bh. We attempt to reproduce the entire line profile at different epochs using a least-squares minimisation {\sc PYTHON} script, which provides a multi-component fit considering a $\chi^2$ close to one\footnote{H${\beta}$ line profiles were not decomposed because of the large contamination of the continuum near the line profile, which may affect the line measurements. Note that the $\chi^2$ strongly depends on the accuracy of the variance estimation.}. Fig. \ref{fig_specdecomp} presents the results of the multi-component fit at some representative epochs: before ($-98.4$ d), during ($-6.4$ d), and after (25.6 d and 165.8 d) the 2015b maximum. The best fit of the H$\alpha$ profiles are obtained using Lorentzian components in emission (a narrow and a broad component at early epochs, and 3 components at late time), and Gaussian components in absorption for the P-Cygni, when they were visible. The velocity estimates for the emission components are derived measuring their FWHM, while those of the absorbing gas shells are estimated from the wavelengths of the P-Cygni minima with respect to the H$\alpha$ rest wavelength. The velocities of the different gas components are listed in Table \ref{table_specpar}, and their evolution is shown in Fig. \ref{fig_specvel}. The velocity uncertainties were estimated with a bootstrap resampling technique, varying randomly the flux of each pixel according to a normal distribution having variance equal to the noise of the continuum. We did this procedure 100 times, and then took the error as the standard deviations of the fit parameters.

At phases $< 30$ d, the best fit of H${\alpha}$ was obtained with two Lorentzian emission components, and one blueshifted absorption component. The FWHM of the narrow H$\alpha$ emission remains nearly constant, with an average value of $\sim$ 1200 \kms, while the broader component has a fast decline from $\sim$ 6000 \kms\ at early phases to $\sim$ 2600 \kms\ at  $-15.4$ d, later on remaining roughly constant. We note that at 25.6 d, the broad component reaches a velocity of $\sim$ 3950 \kms. This measurement is affected by some uncertainty and cannot be confirmed by a spectrum taken in the following days.

After a careful analysis of the H$\alpha$ line profiles, we are able to distinguish a shallow absorption feature (hereafter labelled P-Cy1) in the blue wing of H$\alpha$, visible from day $-103.3$ to day $-15.4$, thus before the 2015b event. The P-Cy1 absorption is blueshifted by a constant amount of $\sim$ 750 \kms. At the time of the 2015b event, this absorption is no longer visible. The increased temperature allows the formation of features such as the He\,{\sc i} $\lambda$6678 line, as well as other He\,{\sc i} lines. When the temperature falls (at phases $>$ 15 d), the P-Cy1 feature becomes detectable again at an unchanged velocity. At the same time, a second absorption (hereafter labelled as P-Cy2) is observable, blueshifted by  $\sim$ 2100 \kms. 

After the seasonal gap, at later phases ($>130$ d), the H${\alpha}$ profile is well reproduced with three Lorentzian components, which we will call blue, core, and red components, following the labelling in \citet{benetti16} for the Type II-L SN~1996al. The blue and core components are centered at an average wavelength of 6522 $\AA$ and 6563 $\AA$, respectively. Instead, the red component displays a slight evolution from 6577 $\AA$ (at 132.5 d) to 6587 $\AA$ (at 286.6 d). The FWHM of these components shows different behaviour: the blue and red components start from different values, but after 190 d converge to similar and relatively constant widths of 1250/1100 \kms. The width of the core component, instead, experiences a slow increase the same time interval from $\sim$1050 \kms\ to $\sim$ 1500 \kms. A small absorption can also be noticed on the top of the H${\alpha}$ profile's blue component in the first spectrum obtained after the transient's disappearance behind the Sun. This feature may correspond to the P-Cy2 absorption discussed before, with a velocity of $\sim$ 1850 \kms\ (see Fig. \ref{fig_specvel}).

We also estimated the evolution of the total luminosity of the H$\alpha$ line (see bottom panel of Fig. \ref{fig_specvel}, and Table \ref{table_specpar}). As expected, the H$\alpha$ luminosity evolves in a similar fashion as the broadband light curves. It is roughly constant at $\sim$ 1.7 $\times$ 10$^{39}$ erg s$^{-1}$ during the 2015a event, then peaks at 18.0 $\times$ 10$^{39}$ erg s$^{-1}$ in he 2015b maximum, and decreases thereafter. At late phases it remains nearly constant at $\sim$ 2.3 $\times$ 10$^{39}$ erg s$^{-1}$.

\begin{table*}
 \centering
  \caption{Main parameters as inferred from the spectra of SN~2015bh. The velocities are computed from the decomposition of the H${\alpha}$ profile.}
  \label{table_specpar}
  \scalebox{0.85}{
 \begin{tabular}{@{}ccrcccccccc@{}}
\hline
 Date & MJD & Phase & Temperature$^a$ & Radius$^b$ & FWHM$_\mathrm{H\alpha,nar}$ & FWHM$_\mathrm{H\alpha,br}$ & v$_\mathrm{P-Cy1}$ & v$_\mathrm{P-Cy2}$ & Luminosity$_\mathrm{H\alpha}$ & EW$_\mathrm{H\alpha}$$^c$ \\
 & & (days) & (K) & ($\times\, 10^{14}$ cm) & (\kms) & (\kms) & (\kms) & (\kms) & ($\times\ 10^{39}$ erg s$^{-1}$) & $\AA$\\
\hline
20150209 & 57062.97 & $-$103.3 & 8600  & 1.0 (0.1) & 1500 (300) & 6000 (600) & 1000 (200) &  - & 	         1.5 (0.6) & 400 (80) \\	
20150214 & 57067.07 & $-$99.2  & 8100  & 1.3 (0.2) & 900 (300) & 4500 (800) & 600 (400) & - &  	  	 1.5 (0.6) & 300 (70) \\	
20150214 & 57067.88 & $-$98.4  & 8300  & 1.2 (0.2) & 1000 (300)  & 4800 (500) & 700 (350) & - &  	  	 1.6 (0.6) & 300 (70)\\	
20150311 & 57092.37 & $-$73.9  & 8600  & 1.5 (0.2) & 900 (300)   & 3900 (400) &  700 (200)  & - &  	  	 1.5 (0.6) & 300 (60) \\	
20150326 & 57107.95 & $-$58.3  & 8400  & 1.6 (0.2) & 950 (200)   & 3800 (300)  & 700 (200)  & - &  	  	 2.3 (0.6) & 200 (50) \\	
20150327 & 57108.93 & $-$57.3  & 8600  & 1.6 (0.2) & 1200 (400) & 3700 (300) & 700 (400) &  - & 	  	 2.2 (0.6) & 250 (50) \\	
20140424 & 57136.00 & $-$30.3  & 8300  & 2.0 (0.3) & 1000 (300)   & 2900 (300) & 650 (300) & - &  	  	 2.0 (0.6) & 150 (30) \\	
20150508 & 57150.90 & $-$15.4  & 7100  & 2.4 (0.4) & 1400 (400) & 2600 (400) & 700 (200) & 	 - &   	 1.3 (0.5) & 100 (20) \\	
20150516 & 57158.96 & $-$7.3   & 16500 & 2.9 (0.3) & 1000 (300) & 3900 (500) & - & - &		  	 8.5 (1.3) & 70 (20) \\	
20150517 & 57159.90 & $-$6.4   & 15800 & 3.5 (0.3) & 1400 (200)  & 2900 (300)  & - & - &		  	 11.6 (1.8) & 70 (20) \\	
20150520 & 57162.84 & $-$3.4   & 20400 & 2.4 (0.2) & 1400 (300) & 1700 (600) & - & - &		  	 18.1 (2.6) & 70 (20) \\	
20150525 & 57167.91 & 1.6      & 13500 & 5.1 (0.6) & 1150 (300)  & 2800 (300) &  - & 2200 (800) & 	  	 11.1 (1.7) & 50 (10) \\	
20150609 & 57182.89 & 16.6     & 7900  & 11.1 (1.8) & 1200 (400) & 3000 (900) & 800 (300) &1900 (600) & 	 9.8 (1.5) & 60 (10) \\	
20150618 & 57191.90 & 25.6     & 7300  & 11.4 (1.8) & 1100 (200)  & 3900 (400) &  900 (200)  & 2200 (200)  &	 5.9 (1.0) & 50 (10) \\  
\hline
\hline
  & & &  &  & FWHM$_\mathrm{blue}$ & FWHM$_\mathrm{core}$ & FWHM$_\mathrm{red}$ & v$_\mathrm{P-Cy2}$ & &\\
 & &  &  &  & (\kms) & (\kms) & (\kms) & (\kms) &  &\\
 \hline
20151003 & 57298.78 & 132.5  & -  & - & 1500 (300) & 1000 (400) & 650  (300) & 1850 (300) & 2.2 (0.6) & 900 (180) \\
20151105 & 57332.04 & 165.8  & -  & - & 1400 (300) & 1400 (400) & 100  (300) &	  -    	  & 2.1 (0.6) & 950 (190) \\
20151113 & 57339.95 & 173.7  & -  & - & 1300 (300) & 1000 (400) & 800  (300) &	   -   	   &2.3 (0.6) & 1000 (200) \\
20151126 & 57353.13 & 186.9  & -  & - & 1300 (200) & 1300 (300) & 1000  (300) &	   -   	  & 2.5 (0.6) & 1200 (240) \\
20151206 & 57362.45 & 196.2  & -  & - & 1200 (300) & 1300 (300) & 1100 (300) &	    -  	  & 2.1 (0.6) & 1200 (240) \\
20160102 & 57390.25 & 224.0  & -  & - & 1250 (300) & 1400 (300) & 1100 (300) &	    -  	  & 2.7 (0.6) & 1500 (300) \\
20160305 & 57452.89 & 286.6  & -  & - & 1300 (200) & 1500 (300) & 1100 (200) &      -       & 2.2 (0.6) & 1500 (300) \\
\hline
\end{tabular}}
\begin{flushleft}
$^a$ We consider a conservative uncertainty in the temperature of about $\pm$ 500 K.\\
$^b$ We have propagated the uncertainties from the Stefan-Boltzmann equation.\\
$^c$ We consider a conservative uncertainty in the EW of about 20 per cent of the measurements.\\
\end{flushleft}
\end{table*}

%

\subsection{Late-time spectra}\label{latetime}
\begin{figure*}
\centering
\includegraphics[width=1.7\columnwidth]{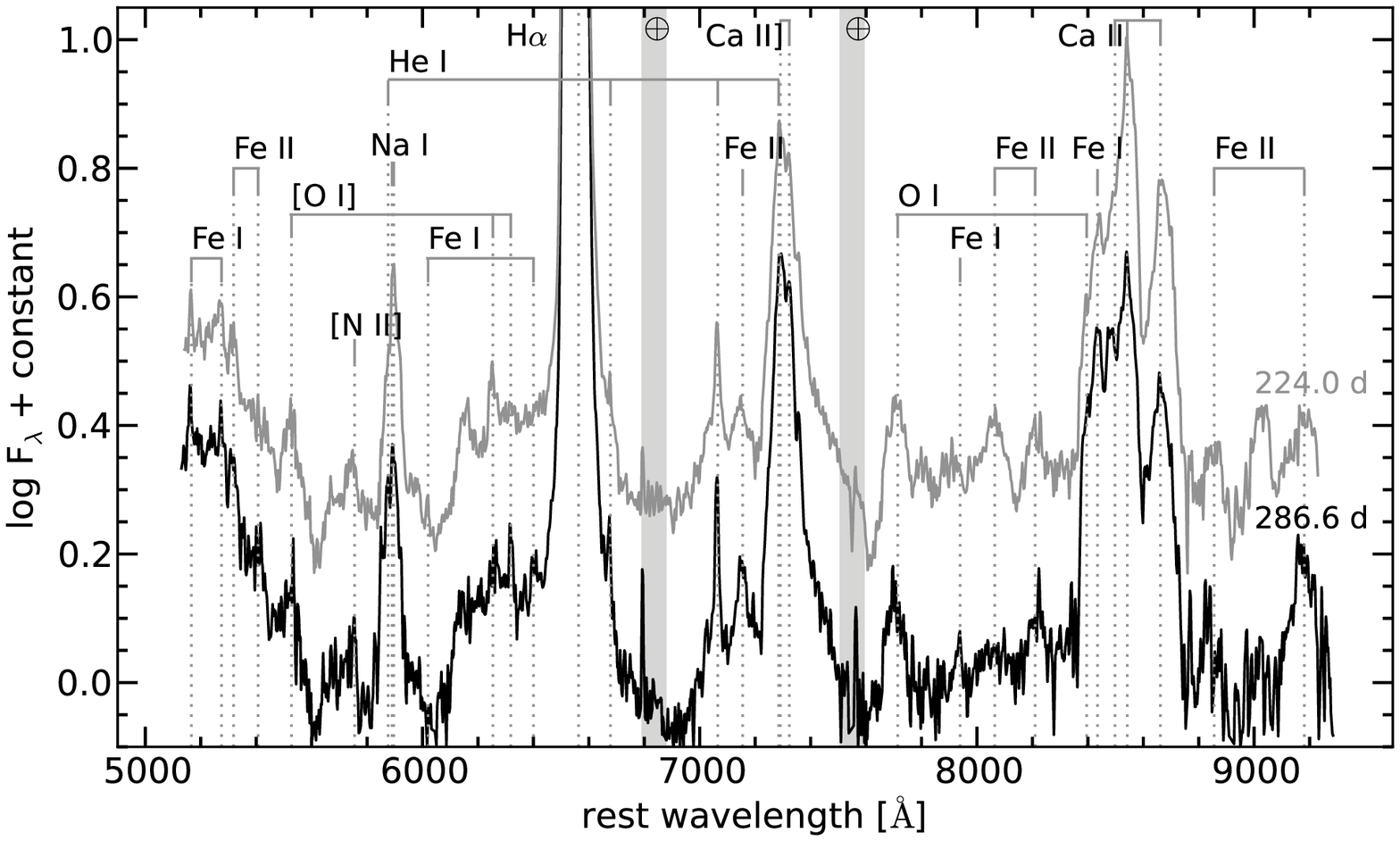}
 \caption{Late-time optical spectra of SN~2015bh at +224.0 ({\it grey line}) and +286.6 d ({\it black line}). The locations of the most prominent spectral features are indicated. A blueshift of 2500 \kms\ has been applied to the vertical lines that mark the rest wavelength of the oxygen.}
\label{fig_latespec}
\end{figure*}

The detailed inspection of the latest spectra of SN~2015bh (Fig. \ref{fig_latespec}) is an excellent opportunity to peer into the very centre of the ejecta star, and constrain the mechanism that gives rise to the 2015a and 2015b events.

As seen before, the late-time spectra still show narrow lines, although combined with by broader features than those seen in the earlier phases. The main change in the spectra is the profile of H$\alpha$. This emission line that dominated the spectra at early time was composed of two components, a narrow feature on top of a broader one. Instead, at late times, the H$\alpha$ profile shows three components, interpreted as the result of the interaction between mostly spherical ejecta with an asymmetric CSM (see section \ref{SNspec_evolhalpha} and \citealt{benetti16}), where the blueshifted H$\alpha$ component arises from faster material than the redshifted one. Note that H$\beta$ is too weak to distinguish this change in the profile. 

Resolved narrow lines of He\,{\sc i} $\lambda$6678, $\lambda$7065, and $\lambda$7283 are also present at their rest position with FWHM around 500 \kms. We do not see any narrow line of He\,{\sc i} $\lambda$5876 in the spectrum at 224.0 d, but instead, we find a broad profile of FWHM $\sim$ 2500 \kms. At 286.6 d, a weak line of He\,{\sc i} appears in the blue side of the profile. In this case, the photons emitted by He\,{\sc i} $\lambda$5876 may be scattered in the optically thick Na\,{\sc iD} lines, resulting in a Na\,{\sc i} doublet feature instead of a He\,{\sc i} feature in the spectra. This mechanism only works if the He\,{\sc i} lines originate in the inner layers of the SN ejecta at these late phases (e.g. see \citealt{benetti16} for more details). 

SN~2015bh late-time spectra also present primordial calcium features (Ca\,{\sc ii}] $\lambda\lambda$7291, 7323, and Ca\,{\sc ii} $\lambda\lambda\lambda$8498, 8542, 8662), similar to core-collapse SNe. We tentatively identify lines of O\,{\sc i} ($\lambda$7774 and $\lambda$8446) and [O\,{\sc i}] ($\lambda$5577 and $\lambda\lambda$6300, 6364), blueshifted by approximately 2500 \kms. This suggests asymmetric SN ejecta, where a possible `blob' of material, which is coming towards us, is being partially ionised by the SN ejecta. Interestingly, the [O I] doublet, is significantly narrower than the permitted oxygen lines (FWHM $\approx$ 1700 km s$^{-1}$ vs. 2800 \kms), hence suggesting that these [O I] lines form in the photoionised CSM. Note that the [O I] doublet is weak and we could have misidentified the feature.

Overall, the spectra of SN~2015bh at late phases are still strongly influenced by the CSM interaction, judging by the presence of narrow emission lines, the pseudo continuum of the iron forest blueward of 5450 $\AA$, as well as of the double-peaked H$\alpha$ profile, and the boxy profile of the Ca\,{\sc ii} NIR triplet. Besides the He\,{\sc i} lines, we can also distinguish other narrow lines at $\sim$ 5166, 5275, 6020, 6400, 7941, and 8439 $\AA$, due to Fe\,{\sc i} from the multiplet 26.

%

\subsection{Spectral comparison}\label{SNspeccomp}

\begin{figure}
\centering
\includegraphics[width=1.\columnwidth]{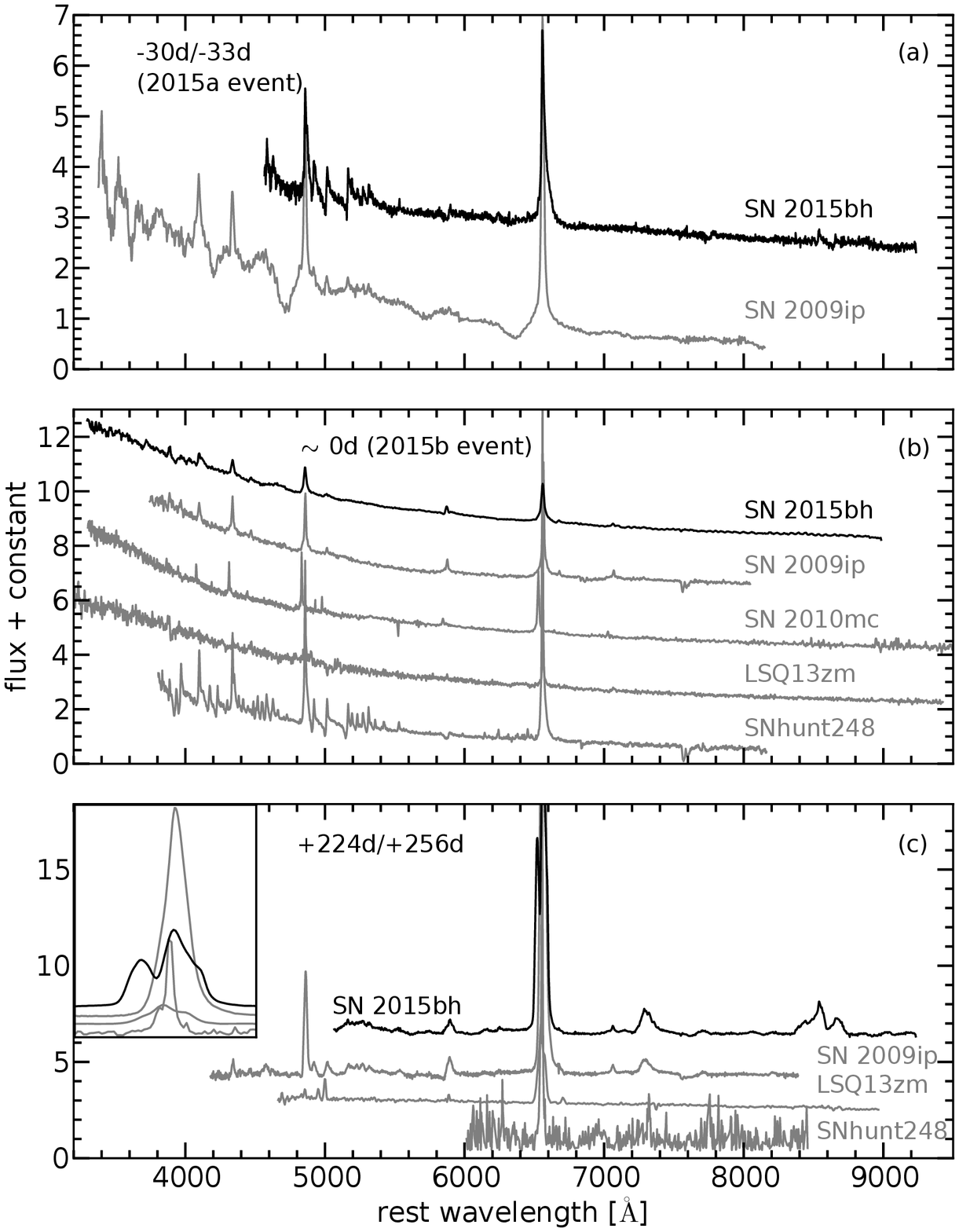}
\caption{Comparison of SN~2015bh spectra during (a) the 2015a event, (b) around the maximum of the major peak, and (c) around 224-256 days after the peak, with those of the transients SNe~2009ip, 2010mc, LSQ13zm, and SNhunt248 at coeval epochs. The H${\alpha}$ profile is blown-up in the insert of panel (c). All spectra have been corrected for their host-galaxy recession velocities and for extinction (values adopted from the literature).  }
\label{fig_speccomp}
\end{figure}

Fig. \ref{fig_speccomp} shows the optical spectra of SN~2015bh at three different epochs, together with the approximately coeval spectra of SN~2009ip \citep{pastorello13,fraser13}, SN~2010mc \citep{ofek13}, LSQ13zm \citep{tartaglia16}, and SNhunt248 \citep{kankare15}\footnote{The spectra are available in the public WISeREP repository \citep{yaron12}.}. The phases of the spectra are relative to their primary maximum (brightest peak of the light curves), and they have been corrected for extinction and redshift using values from the literature.

During the first burst [panel (a) in Fig. \ref{fig_speccomp}], SNe~2015bh and 2009ip show similar narrow features, though SN~2009ip presents a slightly higher temperature of the continuum and broad P-Cygni features associated with the Balmer lines, which are not visible in SN~2015bh. Around the main maximum [panel (b)], all transients of our sample are remarkably similar: this is very likely the phase in which the strength of the ejecta/CSM interaction reaches its peak (cf. section \ref{SNnature}). In passing, we note that the SNhunt248 spectrum at this epoch shows even stronger resemblance to the 2015a rather than the 2015b event. Finally, at late phases [panel (c)], SNe~2015bh and 2009ip show broader features, both in comparison with the other two transients, and with the spectra taken in previous epochs. At these phases, the main difference between these two SNe is the H${\alpha}$ profile [see blow-up window in the panel (c) of Fig. \ref{fig_speccomp}], which in the case of SN~2015bh, is broader and double-peaked, as observed before in interacting SNe such as SN~1996al \citep{benetti16}. Whilst SNhunt248 is considered a SN impostor \citep{mauerhan15,kankare15}, LSQ13zm and SN~2009ip have been proposed to be genuine SNe \citep[e.g.][respectively]{tartaglia16,smith14}.

%

\section{{\sl HST} archival images of SN~2015bh}\label{SNHST}

As described in section \ref{SNph}, the SN~2015bh site was also observed by {\sl HST} with WFPC2 between 2008 and 2009 (WFPC2 Prog.Ids 10877, PI: W. Li, and 11161, PI: A. Soderberg). The field was observed at different times and wavelengths from $F336W$ ($\sim$$U$) to $F814W$ ($\sim$$I$) filters. 
A clear source was detected at the transient position in all the images with $r.m.s.$ uncertainties $<$ 0\parcsec05, through comparison with ground-based, post-discovery NOT+ALFOSC images (taken on 2015 March 27). We performed relative astrometry by geometrically transforming the pre-explosion images to match these post-explosion ones. Assuming these as the deepest images of our collection, we will use them to characterise the nature of SN~2015bh before its discovery. 

\begin{figure}
\centering
\includegraphics[width=1.\columnwidth]{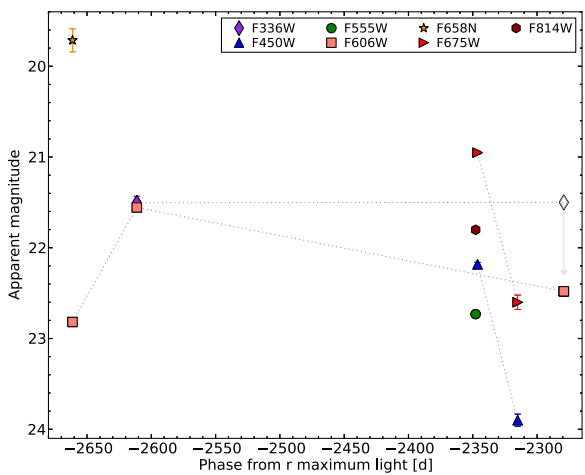}
\caption{Archival {\sl HST} light curves of SN~2015bh. The upper limit is indicated by a symbol with an arrow. The uncertainties for most data points are smaller than the plotted symbols. A colour version of this figure can be found in the online journal. 
}
\label{fig_hstlc}
\end{figure}

\begin{figure}
\centering
\includegraphics[width=1.05\columnwidth]{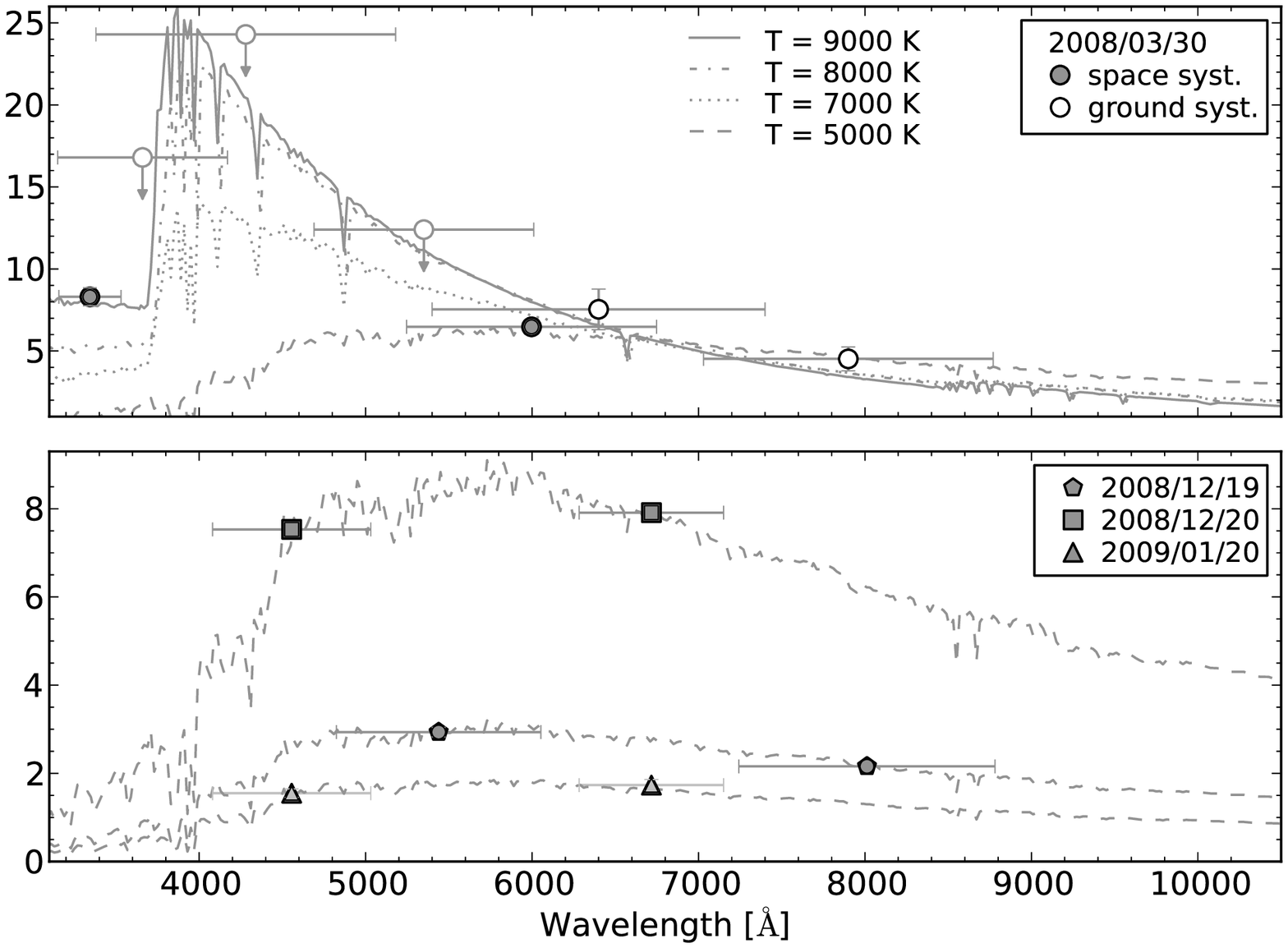}
\caption{Observed spectral energy distribution of the candidate progenitor as measured from multi-epoch images from {\sl HST} ({\it filled symbols}) and ground-based telescopes (when coeval detections were available; {\it empty symbols}). ATLAS synthetic spectra for a star with $T_\mathrm{eff}$ of 9000, 8000, 7000 (log g = 2.0), and 5000 K (log g = 1.5) are also shown. The spectra were obtained assuming solar metallicity. The error-bars along the x-axis match the bandwidths of the corresponding filters.
}
\label{fig_sed}
\end{figure}

During the observation period the star seemed to have some erratic variability (Fig. \ref{fig_hstlc} and Table \ref{table_HSTph}) in a range of $\lesssim$ 1.7 mag. At the same time, other stars (with comparable brightness as our source) observed in the same field remained practically constant. Comparing our SED with the ATLAS synthetic spectra\footnote{\url{http://www.stsci.edu/hst/observatory/crds/castelli\_kurucz\_atlas.html}} of \citet{castelli04}, we approximate the effective temperature of the precursor star for the different {\sl HST} epochs as shown in Fig. \ref{fig_sed} and Table \ref{table_progpar}. We assume near solar metallicity based on the position of the transient in the host galaxy, the assumption of solar metallicity in the centre of NGC~2770, and a metallicity gradient of $-0.06$ dex kpc$^{-1}$ from the nuclear region \citep{thone09}\footnote{For the projected distance from the centre of NGC~2770 to SN~2015bh (2.2 kpc), we estimate 12 + log(O/H) $\approx$ 8.5, which is nearly solar metallicity, following the consideration in \citet{smartt09}.}. Accounting for the extinction and distance modulus reported in section~\ref{SNhostgx}, we also estimate the corresponding luminosity at such epochs (Table \ref{table_progpar}). 

The temperature and luminosity derived for the first epoch (2008 March 30.45 UT) correspond to a massive star of spectral type A, and are consistent with those of massive stars such as LBVs in outburst (e.g. see \citealt{humphreys94} or \citealt{vink12}). These values are confirmed by ground-based observations ($RI$) and limits ($UBV$) obtained with the NOT+ALFOSC on 2008 March 30.89 UTC (Table \ref{table_JCph})\footnote{Note that this last epoch was also used to estimate the temperature of the precursor star on 2008 March 30 (top panel of Fig. \ref{fig_sed}).}. Notice that the temperature at this epoch is mostly determined by the $F336W$ magnitude. The other three {\sl HST} epochs taken 9 months after, instead, indicate a cooler star (even cooler than normal LBVs in eruption), similar to spectral type G. 

The behaviour of the temperature and luminosity is puzzling (see Fig. \ref{fig_hstlc} and Table \ref{table_progpar}). While the irregularity in luminosity is a fair reflection of the variability of this object, the fast subsequent increase in luminosity by a factor of $\sim$3 in one day from 2008 December 19 to 20 is questionable (this would roughly imply that the radius of the star on 2008 December 19 was 70 per cent smaller than a day later).

Giant LBV-like eruptions  (e.g. $\eta$ Car; \citealt{davidson97}), along with major changes in the temperature, are expected to cause variations in the bolometric luminosity \citep{humphreys99}. The high luminosity of the first {\sl HST} epoch of SN~2015bh is consistent with an eruptive state of the transient, while the subsequent {\sl HST} observations show the  progenitor star to have variable luminosity and redder (by 5000 K) colours. These HST observations very likely represent different stages of instability of the star. Notice that this star is always above or close to the Humphreys-Davidson limit \citep{humphreys79}, confirming severe instabilities in the stellar envelope and interior. Unfortunately, the exact time when the eruption occurs or ends is not clear from our measurements. 

Massive stars have been associated to other transients, as is the case of SN~2009ip, where the star was most likely an LBV, with a probable $M_\mathrm{ZAMS}$ of 50-80 $M_{\odot}$ \citep{smith10,foley11}.

\begin{figure*}
\centering
\includegraphics[width=2\columnwidth]{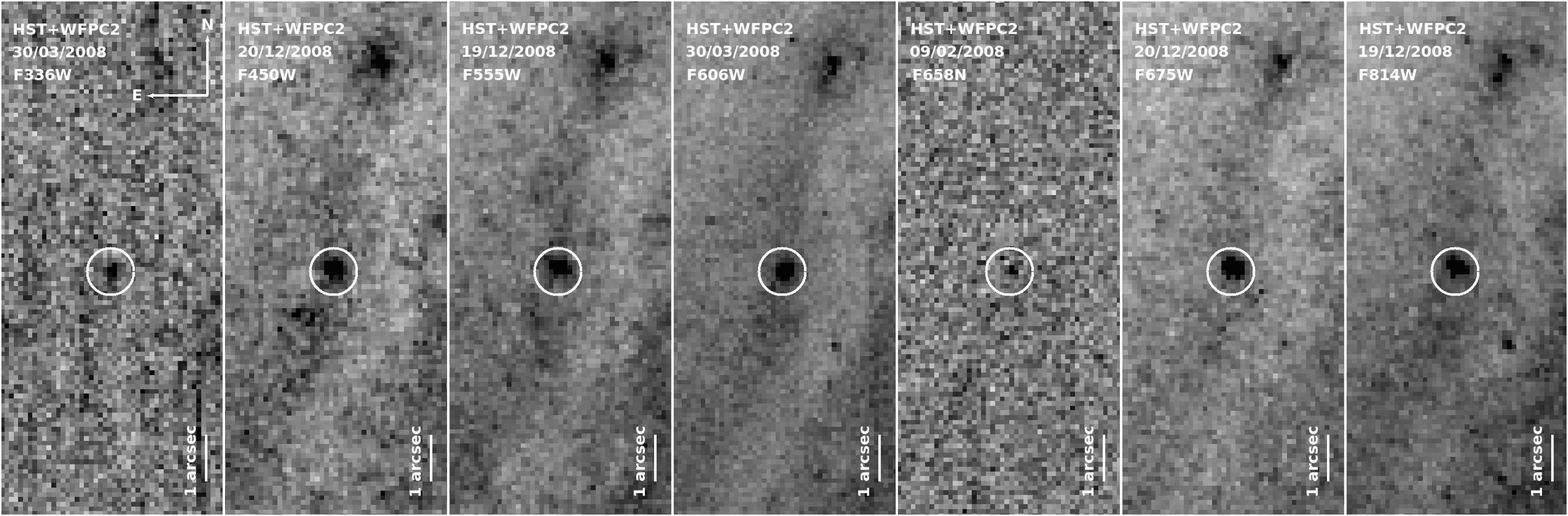}
\caption{Subsections of {\sl HST} WFPC2 images taken between 2008 and 2009 from $F336W$ ({\it first panel on the left}) to $F814W$ ({\it last panel on the right}) filters. The positions of the source at the position of SN~2015bh are indicated by a 5$\sigma$ positional uncertainty circle (0.05 arcsec)}
\label{fig_hstimages}
\end{figure*}

%

\section{On the nature of SN~2015bh}\label{SNnature}

In the previous sections, we analysed the observed properties of SN~2015bh. Combining the information from the light curves, with the evidence of CSM interaction present in the spectra, and the characteristics of the progenitor star, we can attempt to constrain the nature of SN~2015bh. 

As mentioned in the introduction, SN~2009ip has been the benchmark to understand this family of transients. In the past years, several interpretations have been proposed to explain the nature of the most luminous event of SN~2009ip in 2012 (\citealt{pastorello13,soker13,kashi13,mauerhan13a,margutti14,smith14,fraser15,moriya15}). 
Although the non-terminal outburst of a massive star plus subsequent shell-shell collision cannot be ruled out, one of the proposed scenarios for SN~2009ip invokes a faint SN explosion of a compact blue supergiant during the first peak in 2012 (after a series of LBV-like eruptions including the giant eruption in 2009), followed by the interaction of the SN ejecta with a circumstellar shell, producing the 2012b event \citep{mauerhan13a,smith14}. 
Below, we will adopt this scenario in our attempt to explain the chain of events observed in SN~2015bh. 

\begin{figure*}
\centering
\includegraphics[width=1.8\columnwidth]{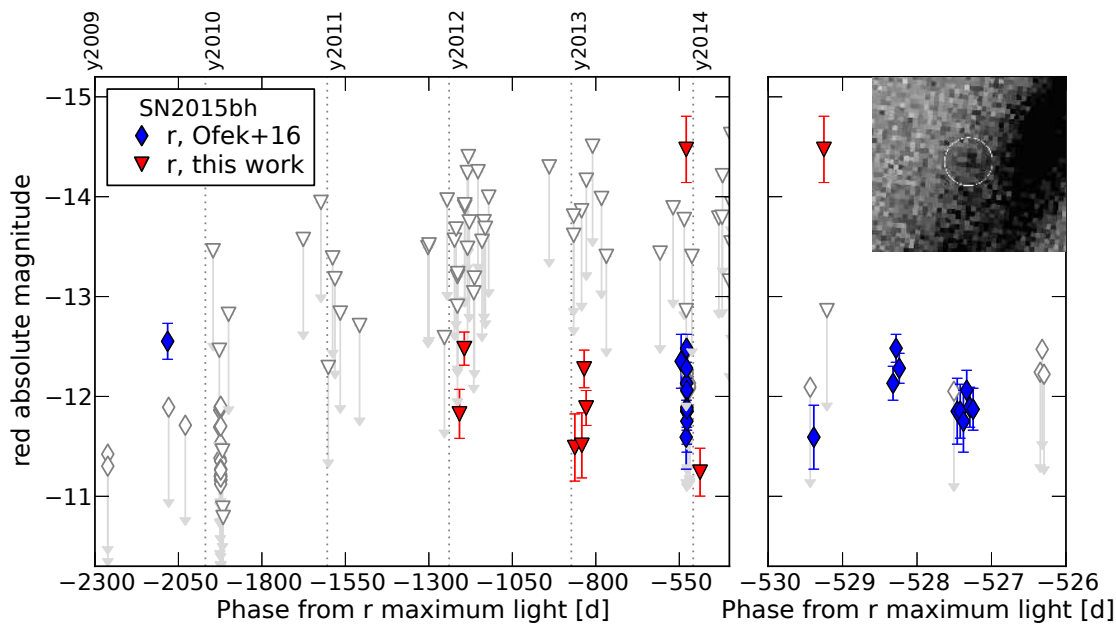} 
\caption{Historical absolute {\sc vegamag} $r$ band light curve of SN~2015bh ({\it filled triangles}) from 2009 to 2014. For comparison we included also the coeval data from \citet[][{\it diamonds}]{ofek16}. Upper limits are indicated by empty symbols with arrows. The right panel shows a zoom of the light curve around the 2013 outburst. The insert is a magnification of the transient position in the image taken on 2013 December 11. A colour version of this figure can be found in the online journal.}
\label{fig_13outburst}
\end{figure*}

\begin{figure}
\centering
\includegraphics[width=1.\columnwidth]{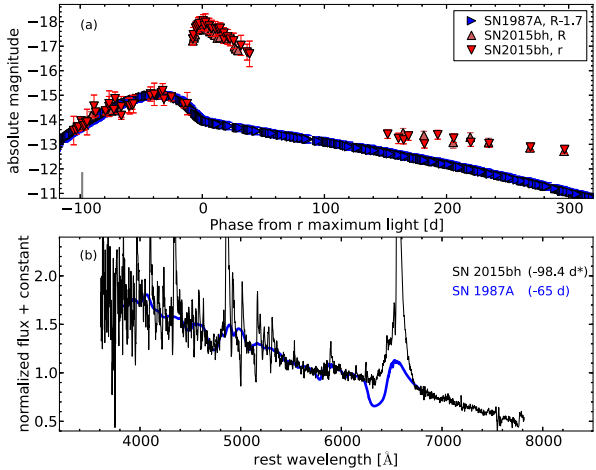}
\caption{Panel (a): Comparison of the absolute {\sc vegamag} $rR$ band light curve of SN~2015bh ({\it filled triangles}), with that of SN~1987A ({\it right rotated triangles}). The SN~1987A light curve has been shifted down by 1.7 mag to match the SN~2015bh light curve at maximum of the 2015a event. The solid mark on the abscissa axis indicate the phase at which SN~2015bh spectrum of panel (b) was obtained. Panel (b): Spectral comparison of SN~2015bh at $-98.4$ d, with that of SN~1987A at coeval epochs. To match the continuum of SN~2015bh, we added a blue  blackbody contribution to the continuum of the SN~1987A spectrum. Both spectra have been corrected for their host-galaxy recession velocities, and normalized to the SN~2015bh spectrum continuum. * Note that the SN~2015bh spectrum is dated at $-98.4$ d from the $r$-band 2015b maximum, and at $-64.5$ d from the $r$-band 2015a maximum. A colour version of this figure can be found in the online journal.
}
\label{fig_87a}
\end{figure}

\begin{enumerate}
\item The first detection of SN~2015bh in archival data dates back to 2002 March 22.89 UTC (MJD = 52355.89), with M$_r$ = $-11.35 \pm$ 0.36 mag [log($L/L_{\odot}$) $\sim$ 6.2]. Later on, the transient was occasionally detected before 2013. As for SN~2009ip, it is likely that SN~2015bh has experienced repetitive outbursts for many years. However, no outbursts have been brighter than M$_r$ = $-12.9$ mag. Properties of these detections are listed in Table \ref{table_progpar}.

It is well known that massive stars occasionally produce giant eruptions during which they can largely increase their luminosity for months to years, and experience major mass loss (e.g. \citealt{smith14c}). Besides, these massive stars are often unstable, possibly alternating between episodes of erratic variability, outbursts and quiescence. Thus, in our case, we may conceive that a shell was expelled around 2002, or even before, and is travelling at approximately 1000 \kms (as derived from both the FWHM of the narrow component of H$\alpha$ and the blueshift of P-Cy1). 

\item At the end of 2013, SN~2015bh experienced an outburst with an absolute magnitude M$_r$ = $-14.6$ mag. This was also detected by \citet{ofek16}, but their detections are around 2 magnitudes fainter (Fig. \ref{fig_13outburst}). As we can see in the insert of the Fig. \ref{fig_13outburst}, there is a detection of a clear source in one amateur image at the transient position on 2013 December 11.03\footnote{Note that this detection is based on a single observation. The field was observed shortly after from another site, and no source brighter than $-13$ mag was detected.}. However, the subsequent decline in a few hours is quite puzzling. A spectrum taken two days after our data, is presented by \citet{ofek16}. This shows evidence for a fast outflow with a velocity of several thousand \kms, similar to that displayed in the SN~2009ip spectra taken one year before the 2012a event (\citealt{pastorello13}). It is noticeable that also this SN~2015bh spectrum was taken around one year before the 2015a event. \citet{pastorello13} suggested that these episodes of ejection of fast material are due to a blast wave probably originated in explosions deeper in the star.
Similar phenomena have been related to LBV eruptions, or even with the Homunculus Nebula surrounding $\eta$ Carinae \citep{smith08}. This blast wave could be the origin of our bright detection (but see also \citealt{soker16}).

As discussed by \citealt{ofek16}, a P-Cygni absorption is also seen during the 2013 outburst spectrum, with a velocity of $\sim$ 1000 \kms. This feature is in agreement with that found in our spectra taken after 2015 February, confirming that material is travelling ahead of that fast outflow. The fact that we do not identify any additional P-Cygni absorptions related with this blast wave at a faster velocity during the 2015a event is however puzzling.

\item At the end of 2014, although we cannot definitely rule out a very massive envelope ejection and no terminal explosion, we favour a scenario where the massive star core-collapsed producing SN~2015bh. This episode has been previously named as the 2015a event. \citet{smith14} favour the scenario of core-collapse SNe from a compact blue supergiants during the faint 2012a event of SN~2009ip. In analogy, the slow rise of the SN~2015bh light curve could be attributed to the explosion of a small initial radius blue progenitor star. Interestingly we find a good agreement between the SN~2015bh and SN~1987A light curves [Fig. \ref{fig_87a}; panel (a)]. However, SN~2015bh is $\sim$ 2 mag fainter compared to SN~1987A, or other normal Type II SNe (typically $< -16$ mag; e.g. \citealt{li11,taddia16}). One possible explanation is that the progenitor of SN~2015bh was a very massive star ($\ge$ 25 M$_{\odot}$) which experienced large fallback of material onto the collapsed core, resulting in a low explosion energy and small amount of ejected $^{56}$Ni (e.g. \citealt{heger03}, or \citealt{moriya10}). We also note that relatively $^{56}$Ni poor SNe II with blue supergiant precursors have been already observed (e.g. SN~2009E; \citealt{pastorello12}).

In addition, we note that the CSM is playing a dominant role in all the SN~2015bh evolution. In the panel (b) of Fig. \ref{fig_87a} we have overplotted an early spectrum of SN~1987A at $\sim -65$ d (manipulated using a blue blackbody  to match the continuum of SN~2015bh), and a spectrum of SN~2015bh at $-98.4$ d\footnote{The SN~2015bh spectrum is dated at $-98.4$ d from the $r$-band 2015b maximum, and at $-64.5$ d from the $r$-band 2015a maximum.}. Although some difference exists in the broad-line velocities, both spectra match surprisingly well, except for the narrow line components visible in SN~2015bh. This indicates that the spectra of SN~2015bh are likely formed by two components: a SN photosphere which radiation is slowly diffused, and a blackbody from the ongoing CSM-ejecta interaction. 

The SN~2015bh spectra during the 2015a event also show a multicomponent H$\alpha$ line (section \ref{SNspec_evolhalpha}), typical of interacting SNe. These  can be explained by radiation coming from different regions of the SN environment. In the case of SN~2015bh, the narrow component is likely due to recombined gas ejected by the star years before, travelling at a velocity $\le$ 1000 \kms. Instead, the broader component (with a velocity of several thousand \kms) provides the velocity of the most recent mass ejection, which decreases with time as the reverse shock propagates into the expanding ejecta. The fact that we can detect this broad H$\alpha$ emission may mean that the cool dense shell formed by the interaction between the SN ejecta and the dense CSM is probably patchy.

\item Shortly after 2015 May 08 (MJD = 57150.90), the newly ejected material collides with a slower and dense CSM, and produces a re-brightening during the 2015b event. This dense material must have been ejected by the star during a recent stellar mass-loss events. However, given the erratic activity of the SN~2015bh progenitor star (see section \ref{SNHST} and also Fig. \ref{fig_abs} and \ref{fig_hstlc}), it is difficult to accurately compute when this gas was expelled. A similar phenomena was also proposed as the cause of the major 2012b peak of SN~2009ip \citep{smith14}. The CSM interaction becomes the primary energy source thanks to the efficient conversion of the kinetic energy into radiation, which is responsible for the increase of the photospheric temperature (see Fig. \ref{fig_spectemp}), the H$\alpha$ luminosity (see panel d of Fig. \ref{fig_specvel}), and the appearance of He\,{\sc i} emission lines.

In order to roughly sketch the opacity of the CSM-interaction region, we derived the total H$\alpha$ emission equivalent width (EW) for SN~2015bh following the analysis of \citealt{smith14}. As we can see in  Fig. \ref{fig_ew}, the EW decreases during the 2015a event to arrived to a minimum during the 2015b peak, to then rise to higher values at late times. We can understand this as the SN ejecta is moving into a denser and thicker CSM during the first phases, to then find a more transparent CSM at late times. The collision with that dense CSM is the cause of the re-brightening during 2015b. The late time EW behaviour of SN~2015bh is consistent with SN~2009ip and other interacting SNe (see \citealt{smith14}), and thus, we expect that the EW of H$\alpha$ can grow in the future.

In this context, the disappearance of the P-Cyg1 absorption remains a puzzle. Perhaps the material is initially photoionized by the SN, which later recombines. Although, this material was expelled in 2002 or earlier, it does not remain unaffected by the hard radiation produced during the 2015b event.

\item Once the shock passes the dense CSM, we see again an absorption at $\sim$ 1000 \kms, along with a second one (P-Cy2) travelling at $\sim$ 2100 \kms. 
If this first absorption originates in the same gas region that produced the similar absorption observed during the 2015a event (P-Cy1), it is hard to explain why it is now detectable again. The new absorption may be part of the un-shocked and relatively dense shell expelled before the SN explosion (which we assume was travelling at a low velocity), and is now shocked by the SN ejecta. This can also be some material ejected after the 2013 outburst (though before the May 2015 event) which, being initially hot and generating wide emission lines, is now cool and detected in absorption, as proposed by \citet{ofek16}. All in all, these blueshifted absorptions indicate that at least two shells or clumps of cooler material move at different velocities along the observer's direction. Evidences of clumpy CSM surrounding very massive stars have been found for LBVs and other supergiant stars. See for example the cases of the progenitors of SNe~1987A \citep{gvaramadze15} and 1996al \citep{benetti16}, but also the Homunculus nebulae in $\eta$ Car \citep{smith12}. However, constraining these asymmetries is a difficult task given the limited available data.

As we can see in Fig. \ref{fig_overplot}, the spectra obtained during the 2015a event (in particular the one at $-15$ d), and that taken at day $+26$ (2015b event) when the brightness of the transient had faded, show very similar features and line velocities. Once again, this could be explained invoking an asymmetric CSM, or  assuming that the CSM is becoming optically thinner after the re-brightening. This claim is also supported by the modest changes in the colour/temperature between the early and late phases. 

\item Finally, at late times, after day +135, the SN ejecta  overtakes the denser CSM region. Broad lines in emission, with the strongest being the NIR Ca\,{\sc ii} feature and Ca\,{\sc ii}] $\lambda\lambda$7291, 7323 (see Fig. \ref{fig_optspec}, and section \ref{latetime}), are detected. Besides, the H$\alpha$ profile changed showing now three components due to the interaction between mostly spherical ejecta with an asymmetric CSM. These spectra are very similar to those of interacting core-collapse SNe such as SN~1996al. The flat light curves suggest that there is still SN ejecta/CSM interaction, preventing the SN following the decline rate predicted by the $^{56}$Co decay (e.g. see Fig. \ref{fig_87a}).

\end{enumerate}

The chain of events of SN~2015bh seems a replica of those observed in SN~2009ip.  
Hence, the similarity between the two transients is remarkable, and can be here summarised:

\noindent
\begin{enumerate}

\item[(1)] Strong evidence of pre-explosion variability or stellar outbursts.

\item[(2)] Faint light curve peak during the first brightening episode.

\item[(3)] Much brighter second peak (episode b), along with strong spectroscopic evidence of ejecta-CSM interaction.

\item[(4)] Very similar late-time spectra, including the possible detection of very weak [O I] lines.

\end{enumerate}

As for SN~2009ip, the interpretation of SN~2015bh is controversial. In this case, we notice a slow rise in the light curve on the 2015a event which closely resembles that of a SN with a blue supergiant progenitor. For this reason, a SN~1987A-like explosion within a H-rich cocoon is a reasonable scenario for SN~2015bh, similar to that proposed for SN~2009ip by \citet{mauerhan13a} and \citet{smith14}. Furthermore, we note that an LBV-like outburst has been proposed to explain the nebula surrounding SK-69202 (the blue supergiant progenitor of SN~1987A; \citealt{smith07}). Both SN~2009ip and SN~2015bh have a faint maximum luminosity of the first peak, which may be indicative that the massive progenitors experienced sub-energetic explosions, facing significant mass fallback onto their stellar cores. Consequently, the amount of ejected $^{56}$Ni is expected to be small. This is consistent with the upper limit of M$_{Ni} \le$ 0.04 M$_{\odot}$ \citep{smith14} measured for SN~2009ip\footnote{\citet{fraser13} estimated M$_{Ni} <$ 0.02 M$_{\odot}$, and  \citet{margutti14} suggested M$_{Ni} <$ 0.08 M$_{\odot}$.}. Large fallback would also explain the lack of prominent [O I] lines in both SNe \citep[but see the discussion in][]{fraser15}.

\citet{ofek16} and \citet{thone16}, still not ruling out alternative scenarios, also seem to favour the final core-collapse for the progenitor of SN~2015bh but during the 2015b event.

Lastly, despite the clues found, the option that the stars did not die still remains a plausible option. In this case, the 2015a+b events can be interpreted as a further (though more severe) mass loss episode plus shell-shell collisions, without leading to a terminal SN explosion. As an additional alternative, the behaviour of SN~2015bh can also be explained in a binary system scenario \citep{soker16}. 
 
\begin{table*}
 \centering
  \caption{Properties of the progenitor star of SN~2015bh.}
  \label{table_progpar}
  {
 \begin{tabular}{@{}ccrcccccl@{}}
\hline
 Date & MJD & Phase & $T$ & $L$ & $E^a$ & FWHM$_\mathrm{H\alpha,nar}$ & FWHM$_\mathrm{H\alpha,br}$ & Note$^b$ \\
 & & (days) & (K) & ($\times\, 10^{39}$ erg s$^{-1}$) & ($\times\, 10^{48}$ erg) & (\kms) & (\kms) & \\
\hline
20020322 & 52355.89 & $-$4810.4 & - & 6 & - & - & - & GBT \\
20080113 & 54478.21 & $-$2688.1 & - & 21 & - & - & - & GBT \\
20080330 & 54555.45 & $-$2610.8 & 9000 & 13 & - & - & - & HST \\
20081219 & 54819.05 & $-$2347.2 & 5000 & 5 & - & - & - & HST \\
20081220 & 54820.51 & $-$2345.8 & 5000 & 15 & - & - & - & HST \\
20090120 & 54851.70 & $-$2314.6 & 5000 & 3 & - & - & - & HST \\
20120215 & 55972.39 & $-$1193.9 & - & 16 & - & - & - & GBT \\
20130111 & 56303.54 & $-$862.7 &  - & 13  & - & - & - & GBT \\
20130208 & 56331.36 & $-$834.9 &  - & 17 & - & - & - & GBT \\
20131211 & 56637.03 & $-$529.3 &  - & 72 & - & - & - & GBT \\
20140121 & 56678.53 & $-$487.8 &  - & 16 & - & - & - & GBT \\
20150420 & 57132.35 & $-$33.9  &  8300 & 140 & 2   & 1150 & 2800 & 2015a event$^c$\\
20150524 & 57166.28 & 0       & 13500 &  2900 & 13 & 1000 & 2900 & 2015b event\\
\hline 
\end{tabular}}
\begin{flushleft}
$^a$ Radiated energy was estimated considering a peak of approximately 140 d durng the 2015a event, and 50 d during the 2015b event.\\
$^b$ GBT = Ground Based Telescope; HST = {\sl Hubble Space Telescope\/}.\\
$^c$ Effective temperature and FWHM velocities were derived from spectra taken at phases -30.3 d and 1.6 d.\\
\end{flushleft}
\end{table*}

\begin{figure}
\centering
\includegraphics[width=1.\columnwidth]{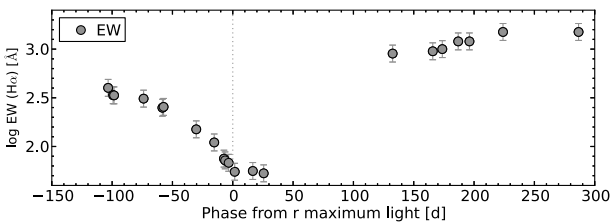}
\caption{Evolution of the total H$\alpha$ emission equivalent width of SN~2015bh. }
\label{fig_ew}
\end{figure}

\begin{figure*}
\centering
\includegraphics[width=1.7\columnwidth]{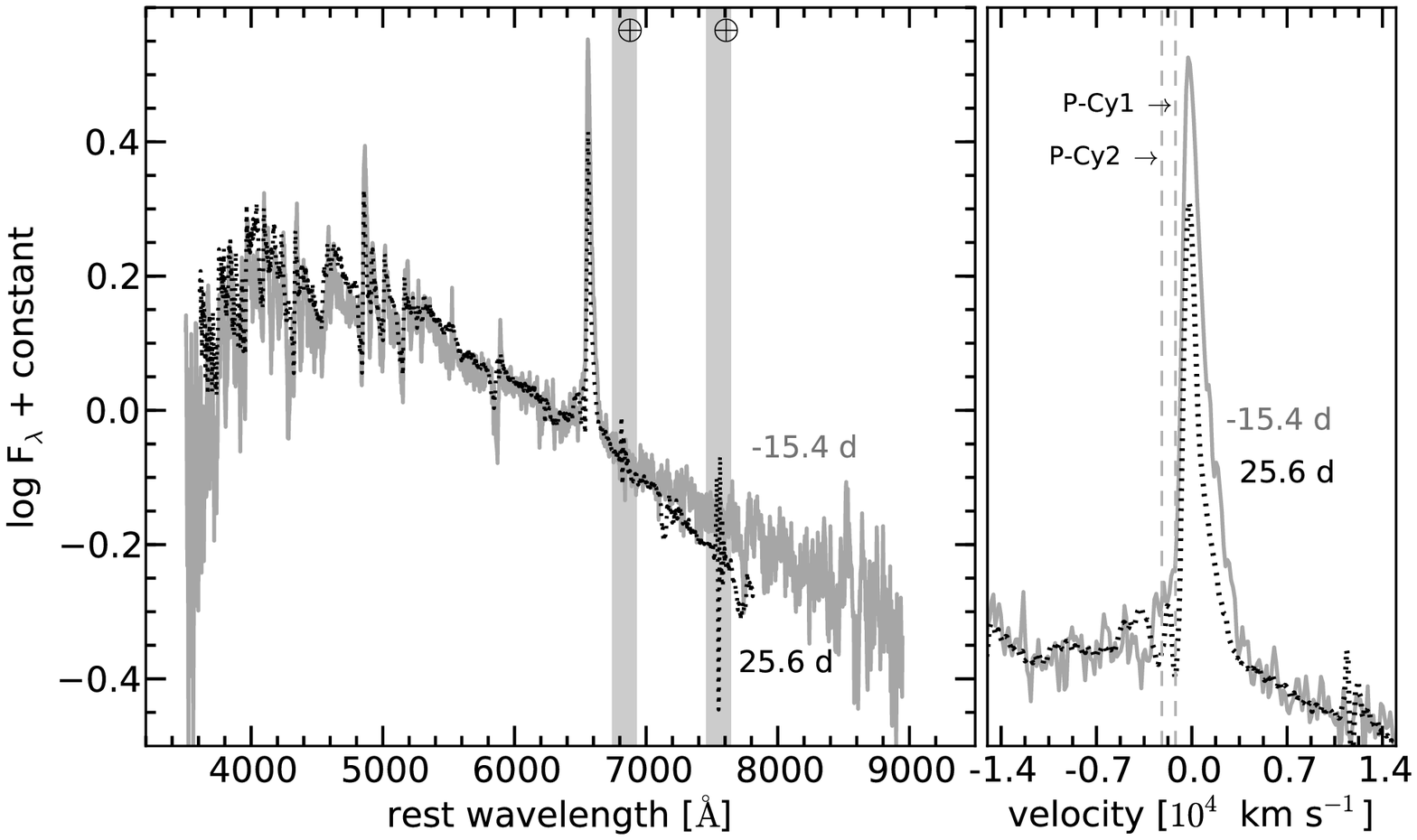}
\caption{Superposition of the 2015 May 08.90 UTC ($-15.4$ d) and 2015 June 18.90 UTC ($25.6$ d) spectra of SN~2015bh. The right panel shows a zoom of the H${\alpha}$ profiles.}
\label{fig_overplot}
\end{figure*}

%

\section{Conclusions}\label{SNconclus}

SN~2015bh was classified as `SN impostor' because of its spectrum, which presents  multi-component H$\alpha$ features reminiscent of those of transients such as SN~2009ip. Although the transient showed a slow evolution in luminosity and a modest variability in its spectral appearance for several months, later on SN~2015bh experienced a major re-brightening during which it increased its luminosity by about 3 mag. 

Analysing in detail the photometric and spectroscopic evolution of SN~2015bh, we have been able to follow a chain of events  similar to that observed in SN~2009ip. Adopting an explosion scenario similar to that proposed  by \citet{mauerhan13a} and \citet{smith14} for SN~2009ip, we propose that the SN~2015bh precursor was likely a massive blue star, possibly similar to that of SN~1987A. However, the progenitor of  SN~2015bh experienced outbursts presumably accompanied by mass loss event along the years. An outburst probably occurred on 2002 (or before), expelling a shell travelling at an average velocity of $\sim$ 1000 \kms. A second one happened at the end of 2013, ejecting material at fast velocity. At the end of 2014, the progenitor star of SN~2015bh possibly explodes experiencing massive fallback of material onto the collapsed core causing a low luminosity explosion. This would be consistent with the low energy of the explosion and the small ejected $^{56}$Ni mass. Later on, fast SN ejecta collide with an outer, dense, and probably non-uniform CSM, producing the re-brightening registered in May 2015. Broad lines of elements such as Ca II and O I, are only detected in the late-time spectra, with the weakening of the ejecta-CSM contaminating contribution. Nonetheless, the CSM interaction which is still affecting the late SN appearance (judging from the flat light curve tail and the presence of boxy and narrow lines in the nebular spectra) prevent us to definitely rule out the non-terminal eruption scenario for SN~2015bh.

One way to shed light on the true nature of SN~2015bh would be continue a photometric and spectroscopic relaxed monitoring to check whether the object vanishes (as expected in the case of a SN), or if another outburst will occur if the progenitor star is still alive, Alongside this, hydrodynamic and/or spectral modeling could bring extra constraints to the nature of the star (stars) that generate SN~2015bh.\\

Erupting massive stars and SN explosions are rare in nearby galaxies. For this reason, the search of these transients is crucial to determine their nature, especially if the link between SN impostors (i.e. luminous pre-SN outbursts) and real supernova explosions is proved. Obviously, multi-wavelength and high-cadence monitoring, along with detailed studies in the X-ray and radio wavelength ranges are crucial for better revealing the structure of the circumstellar environment, and hence to reconstruct the recent stellar mass-loss activity of this type of transients.  

%
%

\section*{Acknowledgments}

We thank S. Leonini for his observations at the Montarrenti Observatory (Siena, Italy), E. E. O. Ishida and U. M. Noebauer for their observations obtained on 2015 March 12, E. Kankare for the SNhunt248 spectra, and A. Harutyunyan for his help.

The research leading to these results has partially been funded by the European Union Seventh Framework Programme (FP7/2007-2013) under grant agreement n. 267251 ``Astronomy Fellowships in Italy'' (AstroFIt)''. NER, AP, SB and LT are partially supported by the PRIN-INAF 2014 (project `Transient Universe: unveiling new types of stellar explosions with PESSTO'). GT and SJS acknowledges European Research Council under the European Union's Seventh Framework Programme (FP7/2007-2013)/ERC Grant agreement n$^{\rm o}$ [291222].  AMG acknowledges financial support by the Spanish Ministerio de Econom\'{i}a y Competitividad (MINECO) grant ESP2013-41268. MF is supported by the European Union FP7 programme through ERC grant number 320360. ST is supported by TRR 33 ``The Dark Universe" of the German Research Foundation. NDR acknowledges postdoctoral support by the University of Toledo and by the Helen Luedtke Brooks Endowed Professorship. The work made use of Swift/UVOT data reduced by P. J. Brown and released in the Swift Optical/Ultraviolet Supernova Archive (SOUSA). SOUSA is supported by NASA's Astrophysics Data Analysis Program through grant NNX13AF35G.

This research is based on observations made with: 
the Nordic Optical Telescope, operated by the Nordic Optical Telescope Scientific Association at the Observatorio del Roque de los Muchachos, La Palma, Spain, of the Instituto de Astrof\'{i}sica de Canarias;
the Gran Telescopio Canarias (GTC), installed in the Spanish Observatorio del Roque de los Muchachos of the Instituto de Astrof\'{i}sica de Canarias, on the island of La Palma;
the Italian Telescopio Nazionale Galileo (TNG) operated on the island of La Palma by the Fundación Galileo Galilei of the INAF (Istituto Nazionale di Astrofisica) at the Spanish Observatorio del Roque de los Muchachos of the Instituto de Astrof\'{i}sica de Canarias;
The Liverpool Telescope is operated on the island of La Palma by Liverpool John Moores University in the Spanish Observatorio del Roque de los Muchachos of the Instituto de Astrof\'{i}sica de Canarias with financial support from the UK Science and Technology Facilities Council; 
the 1.82-m Copernico Telescope and the Schmidt 67/92cm of INAF-Asiago Observatory; 
the 1.22 m Galileo Telescope of Dipartimento di Fisica e Astronomia (Universit\`a di Padova) at the Asiago Observatory; 
the Telescopi Joan Or\'o of the Montsec Astronomical Observatory, which is owned by the Generalitat de Catalunya and operated by the Institute for Space Studies of Catalunya (IEEC); 
the 2m HCT, operated by the Indian Institute of Astrophysics.
Observations reported here were also obtained at Haute-Provence Observatory, CNRS, France
This paper includes data gathered with the 6.5 meter Magellan Telescopes located at Las Campanas Observatory, Chile.
This work is based in part on observations obtained at the MDM Observatory, operated by Dartmouth College, Columbia University, Ohio State University, Ohio University, and the University of Michigan. 
Observations reported here were also obtained at the MMT Observatory, a joint facility of the Smithsonian Institution and the University of Arizona.
The Pan-STARRS1 Surveys (PS1) have been made possible through contributions of the Institute for Astronomy, the University of Hawaii, the Pan-STARRS Project Office, the Max-Planck Society and its participating institutes, the Max Planck Institute for Astronomy, Heidelberg and the Max Planck Institute for Extraterrestrial Physics, Garching, The Johns Hopkins University, Durham University, the University of Edinburgh, Queen's University Belfast, the Harvard-Smithsonian Center for Astrophysics, the Las Cumbres Observatory Global Telescope Network Incorporated, the National Central University of Taiwan, the Space Telescope Science Institute, the National Aeronautics and Space Administration under Grant No. NNX08AR22G issued through the Planetary Science Division of the NASA Science Mission Directorate, the National Science Foundation under Grant No. AST-1238877, the University of Maryland, and Eotvos Lorand University (ELTE) and the Los Alamos National Laboratory.
This work was based in part on observations obtained with CPAPIR at the Observatoire du Mont M\'egantic, funded by the Universit\'e de Montr\'eal, Universit\'e Laval, the Natural Sciences and Engineering Research Council of Canada (NSERC), the Fond qu\'eb\'ecois de la recherche sur la Nature et les technologies (FQRNT) and the Canada Economic Development program. These results also made use of the Discovery Channel Telescope at Lowell Observatory. Lowell is a private, non-profit institution dedicated to astrophysical research and public appreciation of astronomy and operates the DCT in partnership with Boston University, the University of Maryland, the University of Toledo, Northern Arizona University and Yale University. The Large Monolithic Imager construction was supported by a grant AST-1005313 from the National Science Foundation.

This paper is also based on observations made with the Swift telescope: we thank their staffs for excellent assistance; 
on data obtained from the Isaac Newton Group Archive which is maintained as part of the CASU Astronomical Data Centre at the Institute of Astronomy, Cambridge;
and on observations made with the NASA/ESA Hubble Space Telescope, obtained from the data archive at the Space Telescope Science Institute. STScI is operated by the Association of Universities for Research in Astronomy, Inc. under NASA contract NAS 5-26555.
This work has made use of the NASA/IPAC Extragalactic Database (NED), which is operated by the Jet Propulsion Laboratory, California Institute of Technology, under contract with NASA.




\bibliographystyle{mnras}
\bibliography{impostors}



\appendix

\section[]{Tables of photometry and spectroscopy of SN~2015bh.}

\begin{table*}
\centering
\caption{Basic information about the telescopes and instruments used (in alphabetical key order).}
\label{table_setup}
\scalebox{0.75}{
\begin{tabular}{@{}lllll@{}}
\hline
Table key & Telescope & Instrument & Pixel-scale & Location \\
& & & (arcsec pixel$^{-1}$) & \\
\hline
AFOSC & 1.82 m Copernico Telescope & AFOSC & 0.52  & Mount Ekar Obs., Asiago, Italy\\
AGBX & 1.00 m Jacobus Kapteyn Telescope & Acquisition and Guidance Unit & 0.31 & Roque de Los Muchachos Obs., La Palma, Spain\\
ALFOSC & 2.56 m Nordic Optical Telescope & ALFOSC  & 0.19  &  Roque de Los Muchachos Obs., La Palma, Spain\\
ANDOR & 0.80 m Cassegrain Telescope & Andor DW436 CCD & 0.77  & Haute-Provence Obs., Alpes-de-Haute-Provence, France\\
B$\&$C & 1.22 m Galileo Telescope & B$\&$C & -  & Mount Pennar Obs., Asiago, Italy\\
CAFOS & 2.20 m Calar Alto Telescope & CAFOS & 0.53  & Calar Alto Obs., Almer\'{i}a, Spain\\
CPAPIR & 1.60 m Ritchey-Chretien Telescope & CPAPIR & 0.89 & Observatoire du Mont M\'egantic, Qu\'ebec, Canada\\
FORS2 & 8.20 m Very Large Telescope-UT1 & FORS2 & 0.13 & European Southern Obs., Cerro Paranal, Chile \\
GPC1 & 1.80 m Pan-STARRS Telescope 1 & GPC1 & 0.25  & Mount Haleakala Obs., Maui, USA \\
HFOSC & 2.00 m Himalayan Chandra Telescope & HFOSC & 0.30 & Indian Astronomical Obs., Hanle, India \\
HST & 2.40 m Hubble Space Telescope & WFPC2 & 0.05/0.10$^a$ & - \\
IMACS & 6.50 m Magellan Telescope & IMACS & 0.11 & Las Campanas Observatory, Chile\\
IO:O & 2.00 m Liverpool Telescope & IO:O   & 0.30  & Roque de Los Muchachos Obs., La Palma, Spain\\
ISIS & 4.20 m William Hershel Telescope & ISIS & - & Roque de Los Muchachos Obs., La Palma, Spain\\
LMI & 4.30 m Discovery Channel Telescope & LMI &	0.24 & Lowell Observatory, Happy Jack, AZ, USA \\
LRS & 3.58 m Telescopio Nazionale Galileo & LRS  & 0.25  &  Roque de Los Muchachos Obs., La Palma, Spain\\
MEIA & 0.80 m Joan Or\'o Telescope & MEIA & 0.13  & Montsec Astronomical Obs., Catalunya, Spain\\
MMT & 6.50 m Multiple Mirror Telescope & BlueChannel  & 0.60 & MMT Observatory, Arizona, USA\\
NOTCAM & 2.56 m Nordic Optical Telescope & NOTCAM & 0.24  & Roque de Los Muchachos Obs., La Palma, Spain\\
OSIRIS & 10.40 m Gran Telescopio CANARIAS & OSIRIS  & 0.25  &  Roque de Los Muchachos Obs., La Palma, Spain\\
OSMOS & 2.40 m Hiltner Telescope & OSMOS & 0.27 & Michigan-Dartmouth-MIT Obs., Arizona, USA\\
PFCU & 2.54 m Isaac Newton Telescope & Prime Focus Cone Unit & 0.59 & Roque de Los Muchachos Obs., La Palma, Spain\\
PRIME & 4.20 m William Herschel Telescope & Prime Imaging & 0.42 & Roque de Los Muchachos Obs., La Palma, Spain\\
RATCam & 2.00 m Liverpool Telescope & RATCam & 0.28  & Roque de Los Muchachos Obs., La Palma, Spain\\
SDSS & 2.50 m Telescope & Sloan Digital Sky Survey CCD & 0.39  & Apache Point Obs., New Mexico, USA\\
SWIFT & 0.30 m Ritchey-Chretien UV/optical Telescope & SWIFT & 0.50 & - \\
WFC & 2.54 m Isaac Newton Telescope & Wide Field Camera & 0.33  & Roque de Los Muchachos Obs., La Palma, Spain\\
\hline
CAO & 0.36 m Celestron C-14 Telescope & Apogee AP7 CCD & 1.27  & Coddenham Astronomical Obs., Coddenham, U.K.\\ 
CO & 0.30 m Maioni Telescope & SBIG ST-8 & 2.10  & Cortina Obs., Italy\\ 
GO & 0.25 m Newtonian Telescope & DSI-PRO & 1.23  & Gavena Obs., Firenze, Italy \\ 
IAO & 0.60 m reflector telescope & KAF-1001E & 1.45 & Itagaki Astronomical Obs., Teppo-cho, Japan \\ 
MAO & 0.50 m Newtonian Telescope & FLI Proline 4710 CCD & 2.32/1.16$^b$  & Monte Agliale Obs., Lucca, Italy\\ 
MMAO & 0.36 m Celestron C-14 Telescope & Starlight MX 916 & 1.45 & Monte Maggiore Astronomical Obs., Forl\'{i}, Italy \\ 
MO & 0.53 m Ritchey-Chr\'etien Telescope & Apogee Alta U4000 & 1.11 & Montarrenti Obs., Siena, Italy \\ 
OO & 0.40 m Dal Kirkam Telescope & DTA-Electra CCD & 1.17   & Orciatico Obs., Pisa, Italy \\ 
PO & 0.36 m Celestron C-14 Telescope & Starlight SXV-H9 & 1.27  & Pennell Obs., South Wonston, U.K. \\ 
\hline
\end{tabular}
}
\begin{flushleft}
$^a$ WFPC2 contains four chips. SN~2015bh field was observed with the WF4 chip (0.10 arcsec px$^{-1}$) in some epochs, and with the Planetary Camera (0.05 arcsec px$^{-1}$) in others. \\
$^b$ The image taken on 20120116 was done with binning 1x1 and so, the pixel-scale is 1.16 arcsec pixel$^{-1}$\\
\end{flushleft}
\end{table*}

\begin{table*}
\caption{Optical Johnson Cousins photometry of SN~2015bh (Vega magnitudes)}
\label{table_JCph}
\begin{tabular}{@{}ccccccccc@{}}
\hline 
Date & MJD & Phase$^a$ & U & B & V & R  & I & Instrument key \\ 
 &  & (days) & (mag) & (mag) & (mag) & (mag)  & (mag) &  \\ 
\hline 
19940409 & 49451.93 & -7714.3 & - &  $>  21.0 $&  $>  20.1 $&  - &  - & AGBX  \\ 
19960111 & 50093.19 & -7073.1 & - &  - &  - &  - &  $>  19.8 $& AGBX  \\ 
19961013 & 50095.14 & -7071.1 & - &  - &  - &  - &  $>  18.4 $& AGBX  \\ 
19970305 & 50513.00 & -6653.3 & - &  $>  20.0 $&  - &  - &  - & PRIME  \\ 
20080109 & 54474.46 & -2691.8 & $>  19.1 $&  $>  20.2 $&  $>  19.7 $&  $>  19.4 $&  $>  19.3 $& AFOSC  \\ 
20080110 & 54475.94 & -2690.3 & - &  $>  19.3 $&  $>  19.2 $&  $>  18.9 $&  $>  17.5 $& AFOSC  \\ 
20080112 & 54477.24 & -2689.0 & - &  $>  21.8 $&  $>  20.3 $&  - &  - & RATCam  \\ 
20080112 & 54477.54 & -2688.7 & - &  - &  - &  21.09 (0.18) &  21.10 (0.23) & ALFOSC  \\ 
20080112 & 54477.54 & -2688.7 & $>  21.0 $&  $>  22.6 $&  $>  22.8 $&  - &  - & ALFOSC  \\ 
20080113 & 54478.21 & -2688.1 & - &  - &  - &  21.50 (0.27) &  20.77 (0.30) & LRS  \\ 
20080113 & 54478.27 & -2688.0 & - &  - &  - &  20.86 (0.09) &  - & FORS2  \\ 
20080114 & 54479.57 & -2686.7 & - &  - &  - &  - &  21.91 (0.43) & ALFOSC  \\ 
20080114 & 54479.57 & -2686.7 & $>  21.6 $&  $>  22.7 $&  $>  22.8 $&  $>  22.3 $&  - & ALFOSC  \\ 
20080115 & 54480.19 & -2686.1 & - &  $>  21.8 $&  $>  20.8 $&  - &  - & RATCam  \\ 
20080116 & 54481.20 & -2685.1 & - &  $>  20.3 $&  $>  20.2 $&  - &  - & RATCam  \\ 
20080116 & 54481.24 & -2685.0 & - &  - &  - &  $>  21.6 $&  - & FORS2  \\ 
20080117 & 54482.27 & -2684.0 & - &  $>  21.3 $&  $>  20.4 $&  - &  - & RATCam  \\ 
20080118 & 54483.26 & -2683.0 & - &  $>  21.3 $&  $>  20.1 $&  - &  - & RATCam  \\ 
20080120 & 54485.08 & -2681.2 & - &  $>  19.3 $&  $>  18.7 $&  - &  - & RATCam  \\ 
20080125 & 54490.90 & -2675.4 & - &  $>  21.3 $&  $>  20.2 $&  - &  - & RATCam  \\ 
20080128 & 54493.01 & -2673.3 & - &  - &  - &  21.72 (0.46) &  - & CAFOS  \\ 
20080128 & 54493.01 & -2673.3 & $>  21.4 $&  $>  20.0 $&  $>  21.9 $&  - &  $>  21.4 $& CAFOS  \\ 
20080129 & 54494.12 & -2672.2 & - &  $>  21.9 $&  $>  20.9 $&  - &  - & RATCam  \\ 
20080130 & 54495.12 & -2671.2 & - &  $>  20.2 $&  $>  19.3 $&  - &  - & RATCam  \\ 
20080131 & 54497.08 & -2669.2 & - &  $>  22.2 $&  $>  21.6 $&  - &  - & RATCam  \\ 
20080206 & 54502.11 & -2664.2 & - &  $>  22.3 $&  $>  21.6 $&  - &  - & RATCam  \\ 
20080208 & 54504.15 & -2662.1 & - &  $>  21.3 $&  $>  20.2 $&  - &  - & RATCam  \\ 
20080211 & 54507.09 & -2659.2 & - &  - &  - &  $>  20.2 $&  $>  20.4 $& LRS  \\ 
20080211 & 54507.93 & -2658.3 & - &  $>  21.7 $&  - &  - &  - & RATCam  \\ 
20080212 & 54508.10 & -2658.2 & - &  $>  21.1 $&  $>  19.6 $&  $>  19.2 $&  $>  18.7 $& CAFOS  \\ 
20080228 & 54524.99 & -2641.3 & - &  $>  21.2 $&  $>  21.0 $&  - &  - & RATCam  \\ 
20080301 & 54526.97 & -2639.3 & - &  $>  22.6 $&  $>  21.8 $&  - &  - & RATCam  \\ 
20080304 & 54529.96 & -2636.3 & - &  $>  22.5 $&  $>  21.7 $&  - &  - & RATCam  \\ 
20080310 & 54535.94 & -2630.3 & - &  $>  21.1 $&  $>  20.5 $&  - &  - & RATCam  \\ 
20080330 & 54555.89 & -2610.4 & $>  21.0 $&  $>  21.1 $&  $>  21.6 $&  - &  - & ALFOSC  \\ 
20080330 & 54555.89 & -2610.4 & - &  - &  - &  20.95 (0.18) &  20.84 (0.17) & ALFOSC  \\ 
20080415 & 54571.95 & -2594.3 & - &  $>  20.8 $&  $>  20.9 $&  - &  - & RATCam  \\ 
20080421 & 54577.92 & -2588.4 & - &  - &  $>  20.9 $&  - &  - & RATCam  \\ 
20080427 & 54583.04 & -2583.2 & - &  - &  - &  21.62 (0.23) &  - & FORS2  \\ 
20080707 & 54654.76 & -2511.5 & $>  18.4 $&  $>  20.4 $&  $>  19.4 $&  $>  18.9 $&  $>  18.9 $& AFOSC  \\ 
20080708 & 54655.60 & -2510.7 & - &  $>  19.3 $&  $>  20.1 $&  $>  18.9 $&  $>  19.8 $& AFOSC  \\ 
20090221 & 54883.69 & -2282.6 & $>  20.0 $&  $>  20.3 $&  $>  19.4 $&  - &  - & SWIFT  \\ 
20150210 & 57063.90 &  -102.4 & - &  - &  19.75 (0.18) &  - &  - & AFOSC  \\ 
20150211 & 57064.51 &  -101.8 & $>  18.7 $&  $>  19.2 $&  $>  18.2 $&  - &  - & SWIFT  \\ 
20150211 & 57064.95 &  -101.3 & - &  20.27 (0.06) &  19.68 (0.05) &  - &  - & AFOSC  \\ 
20150214 & 57067.83 &   -98.4 & - &  - &  19.59 (0.04) &  18.99 (0.02) &  - & HFOSC  \\ 
20150216 & 57069.82 &   -96.5 & - &  - &  19.64 (0.03) &  19.09 (0.03) &  18.79 (0.04) & HFOSC  \\ 
20150217 & 57070.79 &   -95.5 & - &  20.03 (0.04) &  19.53 (0.05) &  - &  - & AFOSC  \\ 
20150218 & 57071.89 &   -94.4 & $>  19.6 $&  $>  19.9 $&  $>  18.9 $&  - &  - & SWIFT  \\ 
20150222 & 57075.83 &   -90.4 & - &  - &  19.13 (0.04) &  18.68 (0.04) &  - & HFOSC  \\ 
20150223 & 57076.04 &   -90.2 & - &  19.57 (0.02) &  19.15 (0.03) &  18.70 (0.03) &  18.46 (0.04) & ALFOSC  \\ 
20150305 & 57086.97 &   -79.3 & - &  19.57 (0.20) &  - &  - &  - & IO:O  \\ 
20150306 & 57087.74 &   -78.5 & - &  - &  19.01 (0.03) &  18.53 (0.03) &  18.25 (0.03) & HFOSC  \\ 
20150311 & 57092.07 &   -74.2 & - &  19.42 (0.03) &  18.95 (0.03) &  18.50 (0.03) &  18.23 (0.03) & ALFOSC  \\ 
20150311 & 57092.64 &   -73.6 & - &  - &  18.95 (0.03) &  18.48 (0.03) &  18.21 (0.04) & HFOSC  \\ 
20150317 & 57098.79 &   -67.5 & - &  - &  19.04 (0.02) &  18.49 (0.03) &  18.29 (0.03) & HFOSC  \\ 
20150318 & 57099.97 &   -66.3 & - &  19.51 (0.05) &  18.98 (0.04) &  - &  - & AFOSC  \\ 
20150327 & 57108.90 &   -57.4 & - &  19.25 (0.03) &  18.74 (0.03) &  - &  - & ALFOSC  \\ 
20150411 & 57123.90 &   -42.4 & - &  18.79 (0.03) &  18.29 (0.03) &  - &  - & ALFOSC  \\ 
20150428 & 57140.90 &   -25.4 & - &  19.05 (0.04) &  18.34 (0.04) &  - &  - & ALFOSC  \\ 
\hline  
\end{tabular}
\end{table*}

\begin{table*}
\contcaption{}
\begin{tabular}{@{}ccccccccc@{}}
\hline  
Date & MJD & Phase$^a$ & U & B & V & R  & I & Instrument key \\ 
 &  & (days) & (mag) & (mag) & (mag) & (mag)  & (mag) &  \\ 
\hline  
20150508 & 57150.93 &   -15.3 & - &  19.25 (0.03) &  18.52 (0.02) &  - &  - & ALFOSC  \\ 
20150516 & 57158.72 &    -7.6 & 14.92 (0.06) &  16.10 (0.07) &  16.04 (0.09) &  - &  - & SWIFT  \\ 
20150516 & 57158.94 &    -7.3 & 15.23 (0.03) &  16.02 (0.05) &  15.79 (0.04) &  15.65 (0.07) &  15.46 (0.04) & MEIA  \\ 
20150517 & 57159.88 &    -6.4 & - &  15.82 (0.08) &  15.72 (0.11) &  15.53 (0.12) &  - & LRS  \\ 
20150517 & 57159.90 &    -6.4 & 15.04 (0.03) &  15.78 (0.07) &  15.70 (0.04) &  15.46 (0.10) &  15.35 (0.05) & MEIA  \\ 
20150517 & 57159.92 &    -6.4 & 14.64 (0.06) &  15.85 (0.07) &  15.79 (0.08) &  - &  - & SWIFT  \\ 
20150518 & 57160.19 &    -6.1 & 14.63 (0.05) &  15.84 (0.06) &  15.68 (0.07) &  - &  - & SWIFT  \\ 
20150518 & 57160.91 &    -5.4 & - &  15.68 (0.07) &  15.59 (0.07) &  15.39 (0.09) &  - & MEIA  \\ 
20150519 & 57161.66 &    -4.6 & 14.52 (0.05) &  15.61 (0.06) &  15.57 (0.07) &  - &  - & SWIFT  \\ 
20150520 & 57162.23 &    -4.1 & 14.50 (0.05) &  15.59 (0.07) &  15.65 (0.09) &  - &  - & SWIFT  \\ 
20150520 & 57162.85 &    -3.4 & - &  15.61 (0.01) &  15.42 (0.03) &  15.20 (0.03) &  15.05 (0.02) & ANDOR  \\ 
20150520 & 57162.89 &    -3.4 & 14.76 (0.04) &  15.54 (0.06) &  15.38 (0.04) &  15.21 (0.06) &  15.07 (0.05) & MEIA  \\ 
20150521 & 57163.84 &    -2.4 & - &  15.59 (0.05) &  15.40 (0.01) &  15.07 (0.02) &  14.99 (0.04) & ANDOR  \\ 
20150521 & 57163.89 &    -2.4 & 14.36 (0.05) &  15.54 (0.06) &  - &  - &  - & SWIFT  \\ 
20150522 & 57164.19 &    -2.1 & - &  - &  15.40 (0.06) &  - &  - & SWIFT  \\ 
20150522 & 57164.55 &    -1.7 & 14.36 (0.05) &  15.54 (0.06) &  15.49 (0.07) &  - &  - & SWIFT  \\ 
20150522 & 57164.85 &    -1.4 & - &  15.59 (0.03) &  15.39 (0.01) &  15.19 (0.04) &  14.97 (0.02) & ANDOR  \\ 
20150524 & 57166.81 &     0.5 & 14.41 (0.05) &  15.46 (0.06) &  15.35 (0.07) &  - &  - & SWIFT  \\ 
20150524 & 57166.90 &     0.6 & - &  15.46 (0.03) &  15.37 (0.06) &  15.12 (0.05) &  14.98 (0.04) & MEIA  \\ 
20150525 & 57167.48 &     1.2 & 14.47 (0.05) &  15.59 (0.06) &  15.40 (0.07) &  - &  - & SWIFT  \\ 
20150525 & 57167.89 &     1.6 & - &  15.54 (0.02) &  15.33 (0.02) &  15.11 (0.05) &  15.00 (0.03) & ALFOSC  \\ 
20150525 & 57167.90 &     1.6 & 14.71 (0.04) &  - &  15.37 (0.07) &  15.14 (0.03) &  15.00 (0.04) & MEIA  \\ 
20150526 & 57168.04 &     1.8 & 14.52 (0.05) &  - &  - &  - &  - & SWIFT  \\ 
20150526 & 57168.34 &     2.1 & 14.55 (0.05) &  15.58 (0.06) &  15.44 (0.07) &  - &  - & SWIFT  \\ 
20150526 & 57168.90 &     2.6 & 14.88 (0.02) &  15.61 (0.02) &  15.36 (0.02) &  15.13 (0.01) &  15.00 (0.02) & MEIA  \\ 
20150527 & 57169.47 &     3.2 & 14.65 (0.05) &  15.67 (0.06) &  15.40 (0.07) &  - &  - & SWIFT  \\ 
20150527 & 57169.92 &     3.6 & 14.96 (0.04) &  15.63 (0.06) &  15.40 (0.05) &  15.18 (0.04) &  15.03 (0.05) & MEIA  \\ 
20150528 & 57170.24 &     4.0 & 14.71 (0.05) &  15.74 (0.06) &  15.49 (0.07) &  - &  - & SWIFT  \\ 
20150528 & 57170.90 &     4.6 & 15.03 (0.04) &  15.75 (0.06) &  15.40 (0.07) &  15.19 (0.07) &  15.02 (0.03) & MEIA  \\ 
20150530 & 57172.90 &     6.6 & - &  15.87 (0.08) &  15.52 (0.06) &  15.27 (0.06) &  15.15 (0.09) & MEIA  \\ 
20150531 & 57173.90 &     7.6 & 15.27 (0.04) &  15.88 (0.07) &  15.48 (0.05) &  15.25 (0.09) &  15.12 (0.04) & MEIA  \\ 
20150601 & 57174.92 &     8.6 & 15.39 (0.11) &  15.89 (0.05) &  15.50 (0.05) &  15.34 (0.05) &  15.11 (0.04) & MEIA  \\ 
20150602 & 57175.92 &     9.6 & 15.41 (0.12) &  15.95 (0.03) &  15.61 (0.04) &  15.35 (0.05) &  15.13 (0.06) & MEIA  \\ 
20150604 & 57177.00 &    10.7 & - &  16.01 (0.04) &  15.63 (0.03) &  15.42 (0.04) &  15.19 (0.03) & MEIA  \\ 
20150606 & 57179.00 &    12.7 & 15.79 (0.04) &  16.21 (0.02) &  15.71 (0.04) &  15.54 (0.05) &  15.19 (0.04) & MEIA  \\ 
20150608 & 57181.00 &    14.7 & 15.90 (0.15) &  16.27 (0.34) &  - &  15.51 (0.13) &  15.15 (0.16) & MEIA  \\ 
20150611 & 57184.90 &    18.6 & 16.21 (0.03) &  16.50 (0.04) &  15.97 (0.02) &  15.56 (0.04) &  15.38 (0.06) & ALFOSC  \\ 
20150620 & 57193.00 &    26.7 & - &  - &  16.27 (0.03) &  15.88 (0.05) &  15.55 (0.03) & MEIA  \\ 
20150621 & 57194.00 &    27.7 & - &  - &  16.30 (0.05) &  15.97 (0.09) &  15.60 (0.04) & MEIA  \\ 
20150622 & 57195.89 &    29.6 & 17.13 (0.03) &  17.10 (0.03) &  16.39 (0.02) &  16.04 (0.06) &  15.66 (0.04) & ALFOSC  \\ 
20151011 & 57306.17 &   139.9 & $>  21.0 $&  - &  - &  $>  18.2 $&  - & ALFOSC  \\ 
20151011 & 57306.17 &   139.9 & - &  21.31 (0.19) &  - &  - &  - & ALFOSC  \\ 
20151023 & 57318.12 &   151.8 & - &  - &  20.38 (0.29) &  - &  - & AFOSC  \\ 
20151105 & 57331.12 &   164.8 & - &  21.48 (0.07) &  20.69 (0.05) &  19.39 (0.09) &  18.99 (0.09) & ALFOSC  \\ 
20151109 & 57335.11 &   168.8 & - &  21.41 (0.22) &  20.41 (0.14) &  - &  - & AFOSC  \\ 
20151122 & 57348.15 &   181.9 & - &  21.52 (0.06) &  20.63 (0.06) &  19.50 (0.06) &  19.23 (0.10) & ALFOSC  \\ 
20151203 & 57359.10 &   192.8 & - &  - &  20.68 (0.28) &  - &  - & AFOSC  \\ 
20151203 & 57359.10 &   192.8 & - &  $>  21.4 $&  - &  - &  - & AFOSC  \\ 
20151216 & 57372.18 &   205.9 & - &  - &  20.88 (0.11) &  19.79 (0.18) &  19.71 (0.18) & ALFOSC  \\ 
20151216 & 57372.18 &   205.9 & - &  $>  20.8 $&  - &  - &  - & ALFOSC  \\ 
20151230 & 57386.13 &   219.8 & - &  21.57 (0.09) &  20.94 (0.07) &  19.57 (0.10) &  19.70 (0.06) & ALFOSC  \\ 
20160114 & 57401.11 &   234.8 & - &  21.63 (0.09) &  20.98 (0.07) &  19.79 (0.09) &  19.80 (0.10) & ALFOSC  \\ 
20160129 & 57417.00 &   250.7 & - &  $>  18.1 $&  $>  17.3 $&  - &  - & ALFOSC  \\ 
20160217 & 57435.05 &   268.8 & - &  21.82 (0.13) &  21.18 (0.07) &  20.01 (0.05) &  20.13 (0.07) & ALFOSC  \\ 
20160315 & 57462.96 &   296.7 & - &  21.89 (0.07) &  21.33 (0.07) &  20.15 (0.07) &  20.44 (0.08) & ALFOSC  \\ 
\hline 
\end{tabular}
\begin{flushleft}
$^a$ Phases are relative to $r$ maximum light, MJD = 57166.28 $\pm$ 0.29.\\ 
\end{flushleft}
\end{table*}

\begin{table*}
\caption{Optical Sloan photometry of SN~2015bh (AB magnitudes)}
\label{table_SLph}
\begin{tabular}{@{}ccccccccc@{}}
\hline  
Date & MJD & Phase$^a$ & u & g & r & i  & z & Instrument key\\ 
 &  & (days) & (mag) & (mag) & (mag) & (mag)  & (mag) &  \\ 
\hline  
19941204 & 49690.14 & -7476.1 & - &  - &  $>  18.1 $&  - &  - & PFCU  \\ 
19951224 & 50075.18 & -7091.1 & - &  - &  $>  20.5 $&  - &  - & PFCU  \\ 
19991205 & 51517.12 & -5649.2 & - &  - &  $>  19.7 $&  - &  - & CAO  \\ 
19991213 & 51525.12 & -5641.2 & - &  - &  $>  18.7 $&  - &  - & CAO  \\ 
20000105 & 51548.03 & -5618.2 & - &  - &  $>  18.9 $&  - &  - & CAO  \\ 
20000221 & 51595.93 & -5570.3 & - &  - &  $>  19.4 $&  - &  - & CAO  \\ 
20000306 & 51609.86 & -5556.4 & - &  - &  $>  19.1 $&  - &  - & CAO  \\ 
20010420 & 52019.99 & -5146.3 & - &  - &  $>  17.9 $&  - &  - & CAO  \\ 
20010508 & 52037.90 & -5128.4 & - &  - &  $>  18.8 $&  - &  - & CAO  \\ 
20010909 & 52161.15 & -5005.1 & - &  - &  $>  18.9 $&  - &  - & CAO  \\ 
20011016 & 52198.19 & -4968.1 & - &  - &  $>  19.6 $&  - &  - & CAO  \\ 
20011029 & 52211.13 & -4955.2 & - &  - &  $>  19.1 $&  - &  - & CAO  \\ 
20011208 & 52251.07 & -4915.2 & - &  - &  $>  19.2 $&  - &  - & CAO  \\ 
20011226 & 52269.99 & -4896.3 & - &  - &  $>  19.9 $&  - &  - & CAO  \\ 
20020125 & 52299.12 & -4867.2 & - &  - &  $>  19.4 $&  - &  - & CAO  \\ 
20020301 & 52334.07 & -4832.2 & - &  - &  $>  19.0 $&  - &  - & CAO  \\ 
20020322 & 52355.89 & -4810.4 & - &  - &  21.55 (0.33) &  - &  - & WFC  \\ 
20020424 & 52388.89 & -4777.4 & - &  - &  $>  19.2 $&  - &  - & CAO  \\ 
20021102 & 52580.11 & -4586.2 & - &  - &  $>  19.1 $&  - &  - & CAO  \\ 
20021231 & 52639.33 & -4526.9 & $>  21.5 $&  $>  22.1 $&  $>  21.7 $&  $>  21.8 $&  $>  20.4 $& SDSS  \\ 
20030105 & 52644.97 & -4521.3 & - &  - &  $>  19.4 $&  - &  - & CAO  \\ 
20030202 & 52672.99 & -4493.3 & - &  - &  $>  19.5 $&  - &  - & CAO  \\ 
20030219 & 52689.05 & -4477.2 & - &  - &  $>  19.4 $&  - &  - & CAO  \\ 
20030312 & 52710.04 & -4456.2 & - &  - &  $>  19.1 $&  - &  - & CAO  \\ 
20030326 & 52724.01 & -4442.3 & - &  - &  $>  18.6 $&  - &  - & CAO  \\ 
20030402 & 52731.93 & -4434.3 & - &  - &  $>  19.6 $&  - &  - & CAO  \\ 
20051105 & 53679.14 & -3487.1 & - &  - &  $>  19.6 $&  - &  - & CAO  \\ 
20060121 & 53756.14 & -3410.1 & - &  - &  $>  18.8 $&  - &  - & CAO  \\ 
20061027 & 54035.81 & -3130.5 & - &  - &  $>  19.2 $&  - &  - & IAO  \\ 
20061126 & 54065.96 & -3100.3 & - &  - &  $>  19.5 $&  - &  - & CAO  \\ 
20070114 & 54114.98 & -3051.3 & - &  - &  $>  19.3 $&  - &  - & CAO  \\ 
20070422 & 54212.88 & -2953.4 & - &  - &  $>  19.9 $&  - &  - & CAO  \\ 
20080112 & 54477.20 & -2689.1 & - &  - &  $>  21.1 $&  $>  20.4 $&  - & RATCam  \\ 
20080115 & 54480.25 & -2686.0 & - &  - &  $>  20.4 $&  $>  20.1 $&  - & RATCam  \\ 
20080116 & 54481.20 & -2685.1 & $>  20.8 $&  - &  $>  19.9 $&  $>  19.9 $&  - & RATCam  \\ 
20080117 & 54482.27 & -2684.0 & $>  20.3 $&  - &  $>  21.1 $&  $>  20.5 $&  - & RATCam  \\ 
20080118 & 54483.27 & -2683.0 & - &  - &  $>  20.7 $&  $>  20.4 $&  - & RATCam  \\ 
20080120 & 54485.09 & -2681.2 & - &  - &  $>  18.8 $&  $>  18.7 $&  - & RATCam  \\ 
20080125 & 54490.92 & -2675.4 & - &  - &  $>  20.1 $&  $>  20.0 $&  - & RATCam  \\ 
20080129 & 54494.13 & -2672.2 & $>  21.7 $&  - &  $>  20.1 $&  $>  19.8 $&  - & RATCam  \\ 
20080130 & 54495.13 & -2671.2 & $>  20.7 $&  - &  $>  19.4 $&  $>  18.5 $&  - & RATCam  \\ 
20080131 & 54497.05 & -2669.2 & $>  21.9 $&  - &  $>  21.4 $&  $>  21.3 $&  - & RATCam  \\ 
20080206 & 54502.12 & -2664.2 & - &  - &  $>  21.9 $&  $>  21.4 $&  - & RATCam  \\ 
20080208 & 54504.17 & -2662.1 & - &  - &  $>  20.1 $&  - &  - & RATCam  \\ 
20080229 & 54525.00 & -2641.3 & - &  - &  $>  21.0 $&  $>  19.8 $&  - & RATCam  \\ 
20080301 & 54526.99 & -2639.3 & - &  - &  $>  22.3 $&  $>  22.4 $&  - & RATCam  \\ 
20080304 & 54529.97 & -2636.3 & - &  - &  $>  21.7 $&  $>  21.8 $&  - & RATCam  \\ 
20080310 & 54535.96 & -2630.3 & - &  - &  $>  20.8 $&  - &  - & RATCam  \\ 
20080415 & 54571.97 & -2594.3 & - &  - &  $>  19.9 $&  - &  - & RATCam  \\ 
20080421 & 54577.93 & -2588.3 & - &  - &  $>  21.7 $&  - &  - & RATCam  \\ 
20081230 & 54830.01 & -2336.3 & - &  - &  $>  19.8 $&  - &  - & MO  \\ 
20100124 & 55220.05 & -1946.2 & - &  - &  $>  19.3 $&  - &  - & MAO  \\ 
20100211 & 55238.43 & -1927.8 & - &  $>  22.0 $&  - &  - &  - & GPC1  \\ 
20100211 & 55238.82 & -1927.5 & - &  - &  $>  20.3 $&  - &  - & CAO  \\ 
20100221 & 55248.29 & -1918.0 & - &  $>  21.3 $&  $>  21.3 $&  $>  20.9 $&  $>  20.5 $& GPC1  \\ 
20100222 & 55249.29 & -1917.0 & - &  $>  21.5 $&  $>  21.9 $&  $>  21.4 $&  $>  21.1 $& GPC1  \\ 
20100223 & 55250.29 & -1916.0 & - &  $>  21.8 $&  $>  22.0 $&  $>  21.4 $&  $>  20.9 $& GPC1  \\ 
20100304 & 55259.45 & -1906.8 & - &  - &  - &  $>  19.9 $&  - & GPC1  \\ 
20100311 & 55266.43 & -1899.8 & - &  - &  $>  19.9 $&  - &  - & GPC1  \\ 
20100317 & 55272.43 & -1893.8 & - &  $>  21.4 $&  - &  - &  - & GPC1  \\ 
\hline  
\end{tabular}
\end{table*}

\begin{table*}
\contcaption{}
\begin{tabular}{@{}ccccccccc@{}}
\hline 
Date & MJD & Phase$^a$ & u & g & r & i  & z & Instrument key\\ 
 &  & (days) & (mag) & (mag) & (mag) & (mag)  & (mag) &  \\ 
\hline 
20101021 & 55490.13 & -1676.2 & - &  - &  $>  19.2 $&  - &  - & CAO  \\ 
20101213 & 55543.17 & -1623.1 & - &  - &  $>  18.8 $&  - &  - & MAO  \\ 
20110104 & 55565.08 & -1601.2 & - &  - &  $>  20.5 $&  - &  - & MAO  \\ 
20110117 & 55578.10 & -1588.2 & - &  - &  $>  19.4 $&  - &  - & MAO  \\ 
20110124 & 55585.15 & -1581.1 & - &  - &  $>  19.6 $&  - &  - & OO  \\ 
20110208 & 55600.93 & -1565.3 & - &  - &  $>  19.9 $&  - &  - & CAO  \\ 
20110329 & 55650.24 & -1516.0 & - &  - &  - &  - &  $>  21.0 $& GPC1  \\ 
20110407 & 55658.87 & -1507.4 & - &  - &  $>  20.0 $&  - &  - & CAO  \\ 
20111029 & 55863.71 & -1302.6 & - &  - &  $>  19.3 $&  - &  - & IAO  \\ 
20111101 & 55866.75 & -1299.5 & - &  - &  $>  19.2 $&  - &  - & IAO  \\ 
20111206 & 55901.65 & -1264.6 & - &  - &  - &  - &  20.31 (0.12) & GPC1  \\ 
20111217 & 55912.97 & -1253.3 & - &  - &  $>  20.2 $&  - &  - & OO  \\ 
20111226 & 55921.09 & -1245.2 & - &  - &  $>  18.8 $&  - &  - & MAO  \\ 
20120117 & 55943.07 & -1223.2 & - &  - &  $>  19.2 $&  - &  - & MAO  \\ 
20120121 & 55947.91 & -1218.4 & - &  - &  $>  19.1 $&  - &  - & GO  \\ 
20120125 & 55951.89 & -1214.4 & - &  - &  $>  19.8 $&  - &  - & OO  \\ 
20120126 & 55952.02 & -1214.3 & - &  - &  $>  19.5 $&  - &  - & MAO  \\ 
20120126 & 55952.82 & -1213.5 & - &  - &  $>  19.5 $&  - &  - & CAO  \\ 
20120201 & 55958.03 & -1208.2 & - &  - &  20.92 (0.19) &  - &  - & WFC  \\ 
20120211 & 55968.37 & -1197.9 & - &  - &  - &  21.37 (0.23) &  - & GPC1  \\ 
20120215 & 55972.39 & -1193.9 & - &  21.27 (0.11) &  20.26 (0.07) &  - &  - & GPC1  \\ 
20120216 & 55973.56 & -1192.7 & - &  - &  $>  18.8 $&  - &  - & IAO  \\ 
20120216 & 55973.89 & -1192.4 & - &  - &  $>  18.8 $&  - &  - & GO  \\ 
20120224 & 55981.53 & -1184.8 & - &  - &  $>  18.5 $&  - &  - & IAO  \\ 
20120225 & 55982.04 & -1184.2 & - &  - &  $>  19.3 $&  - &  - & CAO  \\ 
20120227 & 55984.51 & -1181.8 & - &  - &  $>  18.3 $&  - &  - & IAO  \\ 
20120301 & 55987.59 & -1178.7 & - &  - &  $>  19.0 $&  - &  - & IAO  \\ 
20120314 & 56000.86 & -1165.4 & - &  - &  $>  19.7 $&  - &  - & OO  \\ 
20120316 & 56002.87 & -1163.4 & - &  - &  $>  19.6 $&  - &  - & MAO  \\ 
20120327 & 56013.48 & -1152.8 & - &  - &  $>  18.5 $&  - &  - & IAO  \\ 
20120408 & 56025.49 & -1140.8 & - &  - &  $>  19.2 $&  - &  - & IAO  \\ 
20120414 & 56031.49 & -1134.8 & - &  - &  $>  19.0 $&  - &  - & IAO  \\ 
20120415 & 56032.27 & -1134.0 & - &  20.97 (0.10) &  - &  - &  - & GPC1  \\ 
20120418 & 56035.44 & -1130.8 & - &  - &  $>  19.1 $&  - &  - & IAO  \\ 
20120428 & 56045.47 & -1120.8 & - &  - &  $>  18.7 $&  - &  - & IAO  \\ 
20121026 & 56226.75 &  -939.5 & - &  - &  $>  18.4 $&  - &  - & IAO  \\ 
20121210 & 56271.63 &  -894.7 & - &  - &  - &  - &  $>  20.4 $& GPC1  \\ 
20130107 & 56299.97 &  -866.3 & - &  - &  $>  18.9 $&  - &  - & OO  \\ 
20130108 & 56300.05 &  -866.2 & - &  - &  $>  19.1 $&  - &  - & MAO  \\ 
20130111 & 56303.54 &  -862.7 & - &  - &  21.25 (0.30) &  - &  - & GPC1  \\ 
20130131 & 56323.64 &  -842.6 & - &  - &  $>  18.9 $&  - &  - & IAO  \\ 
20130201 & 56324.44 &  -841.8 & - &  - &  21.23 (0.29) &  - &  - & GPC1  \\ 
20130208 & 56331.36 &  -834.9 & - &  21.15 (0.21) &  20.47 (0.11) &  - &  - & GPC1  \\ 
20130214 & 56337.33 &  -828.9 & - &  21.49 (0.12) &  20.86 (0.09) &  - &  - & GPC1  \\ 
20130214 & 56337.89 &  -828.4 & - &  - &  $>  18.6 $&  - &  - & GO  \\ 
20130305 & 56356.61 &  -809.7 & - &  - &  $>  18.2 $&  - &  - & IAO  \\ 
20130401 & 56383.41 &  -782.9 & - &  - &  $>  18.8 $&  - &  - & IAO  \\ 
20130415 & 56397.83 &  -768.4 & - &  - &  $>  19.3 $&  - &  - & MAO  \\ 
20130923 & 56558.80 &  -607.5 & - &  - &  $>  19.3 $&  - &  - & IAO  \\ 
20131101 & 56597.76 &  -568.5 & - &  - &  $>  18.9 $&  - &  - & IAO  \\ 
20131118 & 56614.57 &  -551.7 & - &  - &  - &  - &  $>  21.3 $& GPC1  \\ 
20131205 & 56631.09 &  -535.2 & - &  - &  $>  19.0 $&  - &  - & MAO  \\ 
20131211 & 56637.03 &  -529.2 & - &  - &  18.27 (0.30) &  - &  - & OO  \\ 
20131211 & 56637.07 &  -529.2 & - &  - &  $>  19.9 $&  - &  - & MAO  \\ 
20131226 & 56652.65 &  -513.6 & - &  - &  - &  - &  $>  20.4 $& GPC1  \\ 
20131228 & 56654.03 &  -512.2 & - &  - &  $>  19.3 $&  - &  - & OO  \\ 
20140121 & 56678.53 &  -487.8 & - &  - &  21.50 (0.19) &  - &  - & GPC1  \\ 
20140318 & 56734.78 &  -431.5 & - &  - &  $>  19.0 $&  - &  - & MAO  \\ 
20140328 & 56744.64 &  -421.6 & - &  - &  $>  18.9 $&  - &  - & IAO  \\ 
20140329 & 56745.80 &  -420.5 & - &  - &  $>  18.5 $&  - &  - & MAO  \\ 
\hline  
\end{tabular}
\end{table*}

\begin{table*}
\contcaption{}
\begin{tabular}{@{}ccccccccc@{}}
\hline 
Date & MJD & Phase$^a$ & u & g & r & i  & z & Instrument key\\ 
 &  & (days) & (mag) & (mag) & (mag) & (mag)  & (mag) &  \\ 
\hline 
20140419 & 56766.90 &  -399.4 & - &  - &  $>  19.6 $&  - &  - & CAO  \\ 
20140423 & 56770.55 &  -395.7 & - &  - &  $>  18.1 $&  - &  - & IAO  \\ 
20140424 & 56771.55 &  -394.7 & - &  - &  $>  19.2 $&  - &  - & IAO  \\ 
20140426 & 56773.88 &  -392.4 & - &  - &  $>  18.8 $&  - &  - & MAO  \\ 
20140506 & 56783.49 &  -382.8 & - &  - &  $>  19.2 $&  - &  - & IAO  \\ 
20140518 & 56795.83 &  -370.4 & - &  - &  $>  19.4 $&  - &  - & MAO  \\ 
20140607 & 56815.27 &  -351.0 & - &  - &  - &  $>  21.1 $&  - & GPC1  \\ 
20140919 & 56919.81 &  -246.5 & - &  - &  $>  18.5 $&  - &  - & IAO  \\ 
20141004 & 56934.16 &  -232.1 & - &  - &  $>  19.3 $&  - &  - & MAO  \\ 
20141009 & 56939.82 &  -226.5 & - &  - &  $>  18.2 $&  - &  - & IAO  \\ 
20141023 & 56953.78 &  -212.5 & - &  - &  $>  18.7 $&  - &  - & IAO  \\ 
20141028 & 56958.12 &  -208.2 & - &  - &  $>  19.6 $&  - &  - & CO  \\ 
20141028 & 56958.13 &  -208.2 & - &  - &  $>  19.0 $&  - &  - & MAO  \\ 
20141030 & 56960.73 &  -205.5 & - &  - &  $>  18.8 $&  - &  - & IAO  \\ 
20141101 & 56962.11 &  -204.2 & - &  - &  $>  19.4 $&  - &  - & MAO  \\ 
20141107 & 56968.72 &  -197.6 & - &  - &  $>  17.9 $&  - &  - & IAO  \\ 
20141212 & 57003.08 &  -163.2 & - &  - &  $>  19.1 $&  - &  - & MAO  \\ 
20141222 & 57013.07 &  -153.2 & - &  - &  $>  20.0 $&  - &  - & MAO  \\ 
20141223 & 57014.98 &  -151.3 & - &  - &  19.21 (0.41) &  - &  - & CO  \\ 
20141227 & 57018.03 &  -148.2 & - &  - &  $>  18.5 $&  - &  - & MAO  \\ 
20150127 & 57049.41 &  -116.9 & - &  - &  - &  19.11 (0.12) &  - & GPC1  \\ 
20150207 & 57060.95 &  -105.3 & - &  - &  19.50 (0.56) &  - &  - & MMAO  \\ 
20150209 & 57062.96 &  -103.3 & 20.85 (0.20) &  19.88 (0.07) &  19.49 (0.07) &  19.58 (0.07) &  - & AFOSC  \\ 
20150210 & 57063.89 &  -102.4 & 20.78 (0.06) &  20.02 (0.05) &  19.36 (0.04) &  19.38 (0.05) &  - & AFOSC  \\ 
20150211 & 57064.93 &  -101.3 & - &  19.89 (0.06) &  19.36 (0.05) &  19.32 (0.06) &  19.21 (0.14) & AFOSC  \\ 
20150213 & 57066.96 &   -99.3 & - &  19.67 (0.02) &  19.10 (0.02) &  19.06 (0.02) &  - & IO:O  \\ 
20150214 & 57067.94 &   -98.3 & - &  19.71 (0.03) &  19.21 (0.01) &  19.16 (0.02) &  - & IO:O  \\ 
20150216 & 57069.01 &   -97.3 & - &  19.89 (0.03) &  19.34 (0.04) &  19.36 (0.04) &  - & IO:O  \\ 
20150216 & 57069.96 &   -96.3 & - &  19.84 (0.02) &  19.30 (0.02) &  19.35 (0.02) &  - & IO:O  \\ 
20150217 & 57070.76 &   -95.5 & 20.36 (0.07) &  19.77 (0.05) &  19.16 (0.04) &  19.24 (0.05) &  - & AFOSC  \\ 
20150218 & 57071.88 &   -94.4 & - &  - &  19.15 (0.50) &  - &  - & MAO  \\ 
20150224 & 57077.58 &   -88.7 & - &  - &  18.51 (0.43) &  - &  - & IAO  \\ 
20150225 & 57078.02 &   -88.3 & - &  19.13 (0.12) &  - &  18.77 (0.12) &  - & IO:O  \\ 
20150225 & 57078.96 &   -87.3 & - &  19.26 (0.03) &  18.75 (0.02) &  18.74 (0.03) &  - & IO:O  \\ 
20150226 & 57079.97 &   -86.3 & - &  19.33 (0.16) &  - &  18.84 (0.12) &  - & IO:O  \\ 
20150305 & 57086.97 &   -79.3 & - &  19.28 (0.07) &  18.70 (0.05) &  18.86 (0.07) &  18.75 (0.05) & IO:O  \\ 
20150308 & 57089.88 &   -76.4 & - &  - &  18.52 (0.38) &  - &  - & PO  \\ 
20150308 & 57089.97 &   -76.3 & - &  - &  18.40 (0.50) &  - &  - & OO  \\ 
20150312 & 57093.85 &   -72.4 & 20.09 (0.07) &  19.16 (0.04) &  18.65 (0.03) &  18.76 (0.04) &  - & AFOSC  \\ 
20150313 & 57094.81 &   -71.5 & - &  - &  18.20 (0.23) &  - &  - & MAO  \\ 
20150318 & 57099.87 &   -66.4 & - &  - &  18.53 (0.40) &  - &  - & MAO  \\ 
20150318 & 57099.95 &   -66.3 & 19.91 (0.08) &  19.14 (0.05) &  18.89 (0.05) &  18.80 (0.05) &  18.68 (0.10) & AFOSC  \\ 
20150324 & 57105.85 &   -60.4 & - &  - &  18.49 (0.25) &  - &  - & PO  \\ 
20150325 & 57106.87 &   -59.4 & - &  - &  18.50 (0.29) &  - &  - & PO  \\ 
20150327 & 57108.92 &   -57.4 & - &  - &  18.48 (0.03) &  18.53 (0.03) &  - & ALFOSC  \\ 
20150328 & 57109.87 &   -56.4 & - &  - &  18.42 (0.33) &  - &  - & MMAO  \\ 
20150411 & 57123.85 &   -42.4 & - &  - &  18.01 (0.25) &  - &  - & MMAO  \\ 
20150411 & 57123.92 &   -42.4 & - &  - &  18.07 (0.03) &  18.05 (0.02) &  - & ALFOSC  \\ 
20150414 & 57126.92 &   -39.4 & - &  - &  18.02 (0.43) &  - &  - & CO  \\ 
20150420 & 57132.88 &   -33.4 & - &  - &  17.96 (0.19) &  - &  - & PO  \\ 
20150421 & 57133.82 &   -32.5 & - &  - &  17.87 (0.24) &  - &  - & MAO  \\ 
20150422 & 57134.84 &   -31.4 & - &  - &  18.22 (0.21) &  - &  - & MAO  \\ 
20150428 & 57140.91 &   -25.4 & - &  18.59 (0.05) &  18.14 (0.03) &  18.10 (0.03) &  18.15 (0.04) & ALFOSC  \\ 
20150508 & 57150.94 &   -15.3 & - &  18.94 (0.02) &  18.41 (0.03) &  18.35 (0.03) &  18.34 (0.04) & ALFOSC  \\ 
20150511 & 57153.88 &   -12.4 & - &  - &  18.40 (0.43) &  - &  - & CO  \\ 
20150516 & 57158.51 &    -7.8 & - &  - &  15.77 (0.19) &  - &  - & IAO  \\ 
20150516 & 57158.93 &    -7.3 & - &  - &  15.60 (0.26) &  - &  - & PO  \\ 
20150518 & 57160.90 &    -5.4 & - &  - &  15.33 (0.21) &  - &  - & PO  \\ 
20150519 & 57161.96 &    -4.3 & - &  - &  15.26 (0.10) &  - &  - & PO  \\ 
20150520 & 57162.90 &    -3.4 & - &  - &  15.20 (0.25) &  - &  - & PO  \\ 
\hline  
\end{tabular}
\end{table*}

\begin{table*}
\contcaption{}
\begin{tabular}{@{}ccccccccc@{}}
\hline 
Date & MJD & Phase$^a$ & u & g & r & i  & z & Instrument key\\ 
 &  & (days) & (mag) & (mag) & (mag) & (mag)  & (mag) &  \\ 
\hline 
20150521 & 57163.91 &    -2.4 & - &  - &  15.14 (0.23) &  - &  - & PO  \\ 
20150524 & 57166.90 &     0.6 & - &  - &  15.03 (0.20) &  - &  - & PO  \\ 
20150525 & 57167.89 &     1.6 & 15.68 (0.03) &  15.37 (0.03) &  15.22 (0.02) &  15.38 (0.03) &  15.49 (0.03) & ALFOSC  \\ 
20150526 & 57168.91 &     2.6 & - &  - &  15.09 (0.31) &  - &  - & PO  \\ 
20150529 & 57171.92 &     5.6 & - &  - &  15.24 (0.28) &  - &  - & PO  \\ 
20150531 & 57173.92 &     7.6 & - &  - &  15.34 (0.29) &  - &  - & PO  \\ 
20150603 & 57176.93 &    10.7 & - &  - &  15.36 (0.27) &  - &  - & PO  \\ 
20150605 & 57178.94 &    12.7 & - &  - &  15.37 (0.17) &  - &  - & PO  \\ 
20150606 & 57179.92 &    13.6 & - &  - &  15.37 (0.12) &  - &  - & PO  \\ 
20150607 & 57180.93 &    14.7 & - &  - &  15.33 (0.16) &  - &  - & PO  \\ 
20150610 & 57183.85 &    17.6 & 16.90 (0.04) &  15.99 (0.04) &  15.79 (0.05) &  15.77 (0.03) &  15.78 (0.06) & AFOSC  \\ 
20150610 & 57183.93 &    17.7 & - &  - &  15.53 (0.17) &  - &  - & PO  \\ 
20150613 & 57186.93 &    20.7 & - &  - &  15.68 (0.26) &  - &  - & PO  \\ 
20150615 & 57188.95 &    22.7 & - &  - &  15.78 (0.22) &  - &  - & PO  \\ 
20150618 & 57191.95 &    25.7 & - &  - &  15.84 (0.27) &  - &  - & PO  \\ 
20150624 & 57197.93 &    31.7 & - &  - &  16.03 (0.38) &  - &  - & PO  \\ 
20150629 & 57202.93 &    36.7 & - &  - &  16.30 (0.29) &  - &  - & PO  \\ 
20150701 & 57204.93 &    38.7 & - &  - &  16.38 (0.50) &  - &  - & PO  \\ 
20151023 & 57318.11 &   151.8 & - &  21.11 (0.17) &  19.65 (0.08) &  19.76 (0.08) &  19.19 (0.26) & AFOSC  \\ 
20151031 & 57326.94 &   160.7 & - &  21.22 (0.22) &  19.82 (0.15) &  19.85 (0.18) &  19.50 (0.42) & LMI  \\ 
20151105 & 57331.14 &   164.9 & - &  20.99 (0.07) &  19.88 (0.06) &  19.71 (0.06) &  19.51 (0.08) & ALFOSC  \\ 
20151109 & 57335.11 &   168.8 & - &  21.12 (0.15) &  19.72 (0.10) &  19.86 (0.10) &  19.34 (0.07) & AFOSC  \\ 
20151122 & 57348.19 &   181.9 & - &  20.97 (0.08) &  20.05 (0.07) &  20.22 (0.08) &  19.73 (0.08) & ALFOSC  \\ 
20151203 & 57359.08 &   192.8 & - &  21.06 (0.23) &  19.67 (0.10) &  20.08 (0.20) &  19.59 (0.09) & AFOSC  \\ 
20151216 & 57372.17 &   205.9 & - &  21.05 (0.09) &  19.81 (0.08) &  20.02 (0.10) &  - & ALFOSC  \\ 
20151216 & 57372.17 &   205.9 & - &  - &  - &  - &  $>  19.4 $& ALFOSC  \\ 
20151230 & 57386.15 &   219.9 & - &  21.19 (0.08) &  19.98 (0.06) &  20.41 (0.09) &  20.02 (0.08) & ALFOSC  \\ 
20160114 & 57401.14 &   234.9 & - &  21.22 (0.07) &  20.03 (0.05) &  20.51 (0.06) &  20.07 (0.10) & ALFOSC  \\ 
20160217 & 57435.07 &   268.8 & - &  21.43 (0.08) &  20.17 (0.03) &  20.75 (0.06) &  20.48 (0.09) & ALFOSC  \\ 
20160315 & 57462.99 &   296.7 & - &  21.54 (0.07) &  20.27 (0.04) &  21.00 (0.07) &  20.70 (0.08) & ALFOSC  \\ 
\hline  
\end{tabular}
\begin{flushleft}
$^a$ Phases are relative to $r$ maximum light, MJD = 57166.28 $\pm$ 0.29.\\ 
\end{flushleft}
\end{table*}

\begin{table*}
\caption{Near infrared photometry of SN~2015bh (Vega magnitudes)}
\label{table_NIRph}
\begin{tabular}{@{}ccccccc@{}}
\hline  
Date & MJD & Phase$^a$ & J & H & K & Instrument key\\ 
 &  & (days) & (mag) & (mag)  & (mag) &  \\ 
\hline  
20150214 & 57067.64 &   -98.6 & 18.30 (0.23) &  - &  17.91 (0.58) & CPAPIR  \\ 
20150303 & 57084.69 &   -81.6 & - &  - &  17.06 (0.29) & CPAPIR  \\ 
20150307 & 57088.14 &   -78.1 & - &  - &  17.36 (0.25) & NOTCAM  \\ 
20150307 & 57088.57 &   -77.7 & - &  - &  17.18 (0.31) & CPAPIR  \\ 
20150520 & 57162.57 &    -3.7 & 14.98 (0.07) &  - &  - & CPAPIR  \\ 
20150521 & 57163.58 &    -2.7 & 14.94 (0.10) &  - &  14.64 (0.06) & CPAPIR  \\ 
20150527 & 57169.57 &     3.3 & - &  - &  14.47 (0.08) & CPAPIR  \\ 
20150529 & 57171.89 &     5.6 & 14.95 (0.26) &  14.75 (0.24) &  14.44 (0.21) & NOTCAM  \\ 
20150604 & 57177.58 &    11.3 & 14.81 (0.10) &  - &  - & CPAPIR  \\ 
20150607 & 57180.57 &    14.3 & - &  - &  14.60 (0.09) & CPAPIR  \\ 
20150927 & 57292.25 &   126.0 & 18.44 (0.36) &  - &  - & NOTCAM  \\ 
20151024 & 57319.92 &   153.6 & 18.72 (0.43) &  - &  - & CPAPIR  \\ 
20151031 & 57326.92 &   160.6 & - &  - &  18.94 (0.62) & CPAPIR  \\ 
\hline  
\end{tabular}
\begin{flushleft}
$^a$ Phases are relative to $r$ maximum light, MJD = 57166.28 $\pm$ 0.29.\\ 
\end{flushleft}
\end{table*}

\begin{table*}
\caption{Ultra-violet photometry of SN~2015bh (Vega magnitudes)}
\label{table_UVph}
\begin{tabular}{@{}ccccccc@{}}
\hline 
Date & MJD & Phase$^a$ & UVW1 & UVM2 & UVW2 & Instrument key\\ 
 &  & (days) & (mag) & (mag) & (mag) &   \\ 
\hline  
20090221 & 54883.69 & $-$2282.6 & $>  20.8 $&  $>  20.4 $&  $>  20.3 $& SWIFT  \\ 
20131210 & 56636.57 &  $-$529.7 & - &  $>  20.6 $&  - & SWIFT  \\ 
20131217 & 56643.23 &  $-$523.0 & - &  $>  20.6 $&  - & SWIFT  \\ 
20131224 & 56650.10 &  $-$516.2 & - &  $>  20.6 $&  - & SWIFT  \\ 
20131231 & 56657.31 &  $-$509.0 & - &  $>  20.6 $&  - & SWIFT  \\ 
20140107 & 56664.29 &  $-$502.0 & - &  $>  20.6 $&  - & SWIFT  \\ 
20140116 & 56673.40 &  $-$492.9 & - &  $>  20.6 $&  - & SWIFT  \\ 
20140121 & 56678.73 &  $-$487.5 & - &  $>  20.7 $&  - & SWIFT  \\ 
20140128 & 56685.44 &  $-$480.8 & - &  $>  20.6 $&  - & SWIFT  \\ 
20140204 & 56692.33 &  $-$474.0 & - &  $>  20.5 $&  - & SWIFT  \\ 
20140211 & 56699.34 &  $-$466.9 & - &  $>  20.6 $&  - & SWIFT  \\ 
20140218 & 56706.37 &  $-$459.9 & - &  $>  20.6 $&  - & SWIFT  \\ 
20140225 & 56713.34 &  $-$452.9 & - &  $>  20.8 $&  - & SWIFT  \\ 
20140304 & 56720.65 &  $-$445.6 & - &  $>  20.6 $&  - & SWIFT  \\ 
20140311 & 56727.16 &  $-$439.1 & - &  $>  20.7 $&  - & SWIFT  \\ 
20140318 & 56734.21 &  $-$432.1 & - &  $>  20.5 $&  - & SWIFT  \\ 
20140325 & 56741.59 &  $-$424.7 & - &  $>  20.8 $&  - & SWIFT  \\ 
20140401 & 56748.61 &  $-$417.7 & - &  $>  20.6 $&  - & SWIFT  \\ 
20140407 & 56754.69 &  $-$411.6 & - &  $>  20.7 $&  - & SWIFT  \\ 
20140415 & 56762.08 &  $-$404.2 & - &  $>  20.6 $&  - & SWIFT  \\ 
20140422 & 56769.39 &  $-$396.9 & - &  $>  20.7 $&  - & SWIFT  \\ 
20140503 & 56780.29 &  $-$386.0 & - &  $>  20.7 $&  - & SWIFT  \\ 
20140507 & 56784.95 &  $-$381.3 & - &  $>  20.6 $&  - & SWIFT  \\ 
20140513 & 56790.89 &  $-$375.4 & - &  $>  20.6 $&  - & SWIFT  \\ 
20140520 & 56797.36 &  $-$368.9 & - &  $>  20.7 $&  - & SWIFT  \\ 
20140527 & 56804.54 &  $-$361.7 & - &  $>  20.6 $&  - & SWIFT  \\ 
20140604 & 56812.17 &  $-$354.1 & - &  $>  20.6 $&  - & SWIFT  \\ 
20150211 & 57064.51 &  $-$101.8 & $>  19.6 $&  $>  19.7 $&  $>  19.2 $& SWIFT  \\ 
20150218 & 57071.90 &   $-$94.4 & $>  20.4 $&  $>  20.4 $&  $>  20.0 $& SWIFT  \\ 
20150516 & 57158.73 &    $-$7.6 & 15.57 (0.08) &  15.15 (0.06) &  15.02 (0.07) & SWIFT  \\ 
20150517 & 57159.92 &    $-$6.4 & 15.17 (0.07) &  14.83 (0.06) &  14.67 (0.06) & SWIFT  \\ 
20150518 & 57160.19 &    $-$6.1 & 15.01 (0.06) &  14.70 (0.07) &  14.63 (0.05) & SWIFT  \\ 
20150519 & 57161.49 &    $-$4.8 & - &  14.62 (0.06) &  - & SWIFT  \\ 
20150519 & 57161.66 &    $-$4.6 & 14.71 (0.06) &  14.64 (0.06) &  14.46 (0.05) & SWIFT  \\ 
20150520 & 57162.23 &    $-$4.1 & 14.77 (0.06) &  14.59 (0.06) &  14.46 (0.06) & SWIFT  \\ 
20150520 & 57162.50 &    $-$3.8 & 14.72 (0.05) &  14.43 (0.06) &  - & SWIFT  \\ 
20150521 & 57163.89 &    $-$2.4 & - &  - &  14.39 (0.05) & SWIFT  \\ 
20150522 & 57164.19 &    $-$2.1 & 14.57 (0.05) &  14.40 (0.05) &  14.43 (0.05) & SWIFT  \\ 
20150522 & 57164.55 &    $-$1.7 & 14.67 (0.05) &  14.57 (0.06) &  14.53 (0.05) & SWIFT  \\ 
20150524 & 57166.82 &     0.5 & 14.82 (0.06) &  14.67 (0.06) &  - & SWIFT  \\ 
20150525 & 57167.24 &     1.0 & - &  - &  14.53 (0.05) & SWIFT  \\ 
20150525 & 57167.48 &     1.2 & 15.07 (0.06) &  14.86 (0.06) &  14.60 (0.06) & SWIFT  \\ 
20150525 & 57167.58 &     1.3 & 14.96 (0.06) &  - &  - & SWIFT  \\ 
20150526 & 57168.34 &     2.1 & 15.28 (0.07) &  15.02 (0.06) &  14.67 (0.06) & SWIFT  \\ 
20150527 & 57169.47 &     3.2 & 15.45 (0.07) &  15.10 (0.06) &  14.87 (0.06) & SWIFT  \\ 
20150528 & 57170.24 &     4.0 & 15.62 (0.07) &  15.24 (0.06) &  14.94 (0.06) & SWIFT  \\ 
\hline 
\end{tabular}
\begin{flushleft}
$^a$ Phases are relative to $r$ maximum light, MJD = 57166.28 $\pm$ 0.29.\\ 
\end{flushleft}
\end{table*}

\begin{table*}
\caption{{\sl HST} photometry of SN~2015bh (Vega magnitudes)}
\label{table_HSTph}
\begin{tabular}{@{}cccccccccc@{}}
\hline 
Date & MJD & Phase$^a$ & F336W & F450W & F555W & F606W	& F658N & F675W & F814W \\
&   & (days) & (mag) & (mag) & (mag)  & (mag) & (mag) & (mag) & (mag)\\ 
 \hline  
20080209 & 54505.50  & $-$2660.8 & - & - & - & 22.82 (0.04) & 19.71 (0.13) &  - &   - \\
20080330 & 54555.45  & $-$2610.8 & 21.50 (0.07) & - &  - &  21.56 (0.02)  & - &  - &  - \\
20081219 & 54819.05  & $-$2347.2 & - &  - &    22.73 (0.04) & - &  - &  - &    21.80 (0.02) \\ 
20081220 & 54820.51  & $-$2345.8 & - &    22.18 (0.02)  &  - &    - &    - &    20.95 (0.02)  &  - \\
20090120 & 54851.68  & $-$2314.6 & - &    23.90 (0.07)  &  - &    - &    - &    22.60 (0.08) &   - \\
20090229 & 54887.57  & $-$2278.7 & $> 21.5$   &  - &    - &    22.48 (0.02)  & - &    - &    - \\
\hline 
\end{tabular}
\begin{flushleft}
$^a$ Phases are relative to $r$ maximum light, MJD = $57166.28 \pm 0.29$.\\
From WFPC2 manual: F336W: WFPC2 U ($\lambda_c$ = 3342 $\AA$); F450W: Wide B ($\lambda_c$ = 4519 $\AA$); F555W: WFPC2 V ($\lambda_c$ = 5398 $\AA$); F606W: Wide V ($\lambda_c$ = 5935 $\AA$); F656N: H$_{\alpha}$ ($\lambda_c$ = 6564 $\AA$); F675W: WFPC2 R ($\lambda_c$ = 6696 $\AA$); F814W: WPFC2 I ($\lambda_c$ = 7921 $\AA$).\\
\end{flushleft}
\end{table*}

\begin{table*}
 \centering
  \caption{Log of spectroscopy observations of SN~2015bh.}
  \label{table_spec}
  \begin{tabular}{@{}ccrllll@{}}
  \hline
  Date & MJD & Phase$^a$ & Instrument key & Grism or grating + slit& Spectral range & Resolution \\
 & & (days) & & & (\AA) & (\AA) \\
  \hline
20150209 & 57062.97  & $-$103.3 & AFOSC & gm4+1{\parcsec}69 & 3400-8200 & 14 \\  
20150214 & 57067.07  & $-$99.2 & ISIS & R300B/R316R+GG495+1{\parcsec}50 & 3200-9100 & 5\\
20150214 & 57067.88  & $-$98.4 & OSIRIS & R1000B+1{\parcsec}00 & 3650-7850 & 7\\
20150311 & 57092.37 & $-$73.9 & MMT & 300   +1{\parcsec}00 (slit) & 3350-8550      & 7          \\
20150326 & 57107.95  & $-$58.3 & OSIRIS & R1000B+1{\parcsec}00 & 3650-7850 & 7 \\
20150327 & 57108.93  & $-$57.3 & ALFOSC & gm4+1{\parcsec}00 & 3400-9000 & 14 \\
20150423 & 57136.00 & $-$30.3 & IMACS  & 300   +0{\parcsec}70 (slit) & 4250-8500      & 4          \\
20150508 & 57150.90  & $-$15.4 & ALFOSC & gm4+1{\parcsec}00 & 3400-9000 & 14 \\
20150516 & 57158.96 & $-$7.3 & IMACS  & 300   +0{\parcsec}70 (slit) & 4250-8500      & 4          \\
20150517 & 57159.90  & $-$6.4 & LRS & LR-B+1{\parcsec}00 & 3400-8000 & 12 \\
20150520 & 57162.84  & $-$3.4 & B\&C & 300+4{\parcsec}00 & 3350-8000 & 11 \\     
20150525 & 57167.91  & 1.6 & ALFOSC & gm4+1{\parcsec}00 & 3300-9000 & 14 \\
20150609 & 57182.89  & 16.6 & LRS & LR-B/LR-R+1{\parcsec}00 & 3300-9300 & 11 \\
20150618 & 57191.90  & 25.6 & OSIRIS & R1000B+1{\parcsec}00 & 3650-7850 & 7 \\
20151003 & 57298.78 & 132.5 & OSMOS       & VPH-R+1{\parcsec}00 (slit) & 4500-9100	  & 4.1 \\
20151105 & 57332.04 & 165.8 & AFOSC & gm4+1{\parcsec}69 & 3400-8250 & 15 \\  
20151113 & 57339.45 & 173.2 & OSMOS       & VPH-R+1{\parcsec}00 (slit) & 4500-9100	  & 4.1 \\
20151126 & 57353.13 & 186.9 & OSIRIS & R1000B+1{\parcsec}00 & 3640-7870 & 7 \\
20151206 & 57362.45 & 196.2 & OSMOS       & VPH-R+1{\parcsec}20 (slit) & 5600-9000      & 5.5        \\
20160102 & 57390.25 & 224.0 & OSIRIS & R1000R+1{\parcsec}00 & 5100-9300 & 8 \\ 
20160305 & 57452.89 & 286.6 & OSIRIS & R1000R+1{\parcsec}00 & 5100-9300 & 8 \\ 
\hline
\end{tabular}
\begin{flushleft}
$^a$ Phases are relative to $r$ maximum light, MJD = $57166.28 \pm 0.29$.
\end{flushleft}
\end{table*}
\onecolumn
\noindent
$^{1}$ INAF - Osservatorio Astronomico di Padova, vicolo dell'Osservatorio 5, Padova I-35122, Italy. \\
$^{2}$ European Organisation for Astronomical Research in the Southern Hemisphere (ESO), Karl-Schwarzschild-Str. 2, D-85748\\
Garching bei M\"unchen, Germany\\
$^{3}$ Max-Planck-Institut f\"ur Astrophysik, Karl-Schwarzschild-Str. 1, D-85748 Garching bei M\"unchen, Germany\\
$^{4}$ Astrophysics Research Centre, School of Mathematics and Physics, Queen's University Belfast, Belfast BT7 1NN, United Kingdom\\
$^{5}$ Institute of Astronomy, University of Cambridge, Madingley Road, Cambridge CB3 0HA, United Kingdom\\
$^{6}$ George P. and Cynthia Woods Mitchell Institute for Fundamental Physics \& Astronomy, Texas A. \& M. University, Department of \\
Physics and Astronomy, 4242 TAMU, College Station, TX 77843, USA\\
$^{7}$ Institut de Ci\`encies de l'Espai (CSIC-IEEC), Campus UAB, Cam\'{\i} de Can Magrans S/N, 08193 Cerdanyola (Barcelona), Spain\\
$^{8}$ Tuorla Observatory, Department of Physics and Astronomy, University of Turku, V\"ais\"al\"antie 20, FI-21500 Piikki\"o, Finland\\
$^{9}$ Ritter Observatory, Department of Physics and Astronomy, The University of Toledo, Toledo, OH 43606-3390, USA\\
$^{10}$ D\'epartement de physique, Universit\'e de Montr\'eal, C.P. 6128,  Succ.~Centre-Ville, Montr\'eal, Qu\'ebec, H3C 3J7, Canada\\
$^{11}$ Center for Interdisciplinary Exploration and Research in Astrophysics (CIERA) and Department of Physics and Astrophysics,\\
Northwestern University, Evanston, IL 60208, USA\\
$^{12}$ Center for Cosmology and Particle Physics, New York University, 4 Washington Place, New York, NY 10003, USA\\
$^{13}$ Institut d'Astrophysique de Paris, CNRS, and Universite Pierre et Marie Curie, 98 bis Boulevard Arago, 75014, Paris, France\\
$^{14}$ Indian Institute of Astrophysics, Koramangala, Bangalore 560 034, India\\
$^{15}$ Pennell Obs., South Wonston, Winchester, HANTS, SO21 3HE, U.K.\\ 
$^{16}$ Dipartimento di Fisica e Astronomia Galileo Galilei, Universit\`a di Padova, Vicolo dell'Osservatorio 3, Padova I-35122, Italy\\
$^{17}$ Coddenham Astronomical Observatory, Peel House, Coddenham, Suffolk, IP6 9QY, United Kingdom \\ 
$^{18}$ Gavena Obs., Firenze, Italy\\
$^{19}$ Orciatico Obs., Pisa, Italy\\
$^{20}$ Astrophysical Institute, Department of Physics and Astronomy, 251B Clippinger Lab, Ohio University, Athens, OH 45701, USA\\
$^{21}$ Osservatorio Astronomico di Monte Agliale, Via Cune Motrone, 55023 Borgo a Mozzano, Lucca, Italy\\ 
$^{22}$ Monte Maggiore Astronomical Obs., Forl\'{i}, Italy \\ 
$^{23}$ Osservatorio Astronomico del Col Drusci\'e, I-32043 Cortina d'Ampezzo, Italy\\ 
$^{24}$ Itagaki Astronomical Observatory, Teppo-cho, Yamagata 990-2492, Japan \\ 
$^{25}$ Harvard-Smithsonian Center for Astrophysics, 60 Garden St., Cambridge, MA 02138, USA\\
$^{26}$ Aryabhatta Research Institute of Observational Sciences, Manora Peak, Nainital 263002, India \\

\bsp	
\label{lastpage}
\end{document}